\documentclass[aip, amsmath,amssymb, onecolumn, preprint,%
]{revtex4-2}
\usepackage{graphicx}
\usepackage{float,rotating}
\usepackage[caption=false]{subfig}
\usepackage{subeqnarray}
\usepackage{bbold}
\usepackage{amsmath, amscd}
\usepackage{verbatim}
\usepackage{amssymb}
\usepackage{mathrsfs}
\usepackage{float}
\usepackage{color}
\usepackage{tikz}
\usetikzlibrary{decorations.pathmorphing, positioning, arrows.meta}
\usepackage [english]{babel}
\usepackage [autostyle, english = american]{csquotes}\MakeOuterQuote{"}
\usepackage[breaklinks=true,colorlinks=true,linkcolor=blue,urlcolor=blue,citecolor=magenta]{hyperref}
\usepackage{mathtools}

\DeclarePairedDelimiter\ket{\lvert}{\rangle}

\makeatletter
\def\@email#1#2{%
	\endgroup
	\patchcmd{\titleblock@produce}
	{\frontmatter@RRAPformat}
	{\frontmatter@RRAPformat{\produce@RRAP{*#1\href{mailto:#2}{#2}}}\frontmatter@RRAPformat}
	{}{}
}%
\makeatother
\begin{document}
	
	
	\title[]{A Closed-Form Approach to Oscillatory Integrals in Level-Crossing Physics}
	\author{Maseim B. Kenmoe}
	\affiliation{Mesoscopic and Multilayer Structures Laboratory, Faculty of Science, Department of Physics, University of Dschang, Cameroon}
	\email{kenmoe@aims.edu.gh}
	\author{Anicet D. Kammogne}
	\affiliation{Mesoscopic and Multilayer Structures Laboratory, Faculty of Science, Department of Physics, University of Dschang, Cameroon}%
		
	\date{\today}
	
	\begin{abstract}
		We present a closed-form, exact analytical solution, valid at finite times, to a class of multiple integrals with highly oscillatory kernels. Our approach leverages the intimate connection between these integrals and the minimal level-crossing model, namely the Landau-Zener model. Benchmarking against data from numerical simulations demonstrates excellent agreement validating our analytical method. Impacts of our results in level-crossing dynamics are also discussed. A dedicated Mathematica package named {\bf OscillatoryIntegralAnalytical.wl} publicly accessible in our \href{https://github.com/Kenmax15/Closed-Form-Approach-to-Oscillatory-Integrals/tree/main}{GitHub repository} allows generating the integrals for arbitrary order. 
	\end{abstract}
	
	\maketitle
	
	\section{Introduction}\label{Sec1}
	Oscillatory integrals are ubiquitous in Mathematical analysis and contemporary Physics\cite{Kayanuma1984, Kayanuma1985, Kayanuma1998, Demkov2001, Volkov2004,Saito2007, Rojo2010, Glasbrenner2023}. They play pivotal role in non-stationary quantum mechanics particularly in quantum interference stemming from coherent superposition (splitting and recombination) of wave functions at level-crossings\cite{Shevchenko2010, Kiselev2013, Kenmoe2017}. They occur in a wide range of scientific disciplines ranging from path integrals theory\cite{Job2025}, diffraction theory\cite{Diffraction, Moshinsky}, financial economics\cite{Sornette2014}, engineering\cite{Shulin2023, Bounds2024} and evolutionary theory\cite{Baake1997, Dziarmaga2005}. 
	A remarkable class of such integrals is exemplified by\cite{Kenmoe2013, Kholodenko2012}
	\begin{eqnarray}\label{equ1}
	\nonumber	\mathcal{I}_{k}(\tau)=\int^\tau_{-\infty}d\tau_{1}\int^{\tau_{1}}_{-\infty}d\tau_{2} ...\int^{\tau_{2k-2}}_{-\infty}d\tau_{2k-1}\int^{\tau_{2k-1}}_{-\infty}d\tau_{2k}\cos[\tau^{2}_{1}-\tau^{2}_{2}]\times\cdots\times\cos[ \tau^{2}_{2k-1}-\tau^{2}_{2k}],  \quad k\ge 1.\hspace{-1cm}\\
	\end{eqnarray}
	Here, $\tau$ denotes time and may also be the arc-length of a cornu spiral, temperature, chemical potential, spatial coordinates or flux. This is a fundamental problem in Mathematics and has so far attracted considerable efforts\cite{Siladadze2013, Job2023}. Except for $\mathcal{I}_{1}(\tau)$, no existing Mathematical method, to the best of our knowledge, allows direct calculation of such integrals at finite times for $k>1$. Instead, they can typically be approximated by using techniques such as the method of steepest descent\cite{Erdelyi1956}, the Levin method\cite{Chen2024}, the Filon-type method\cite{Liu2017}, and the asymptotic expansion method\cite{Jing2012} among others. However, because the kernel rapidly oscillates as $\tau$ increases, the resulting solution often diverges when $\tau\to\infty$. 
	
	On the other hand, this problem intimately connects with the minimal level-crossing model: the two-level Landau-Zener (LZ) problem\cite{Landau1932, Zener1932, Stuckelberg1932, Majorana1932}. The later describes a system  undergoing non-adiabatic transitions at a level-crossing by linear variation of a control parameter. It is usually addressed from both Schr\"odinger and Bloch pictures where it admits exact solutions at both finite- and infinite- times $\tau=\infty$. The connection between LZ and the dynamics of a sphere rolling on a cornu spiral of arc-length $s$ is established in Refs.\onlinecite{Rojo2010a, Rojo2010b} and extensively exploited in Ref.\onlinecite{Kholodenko2012} to solve \eqref{equ1} indirectly at $\tau=\infty$ yielding (see also Ref.\onlinecite{Kenmoe2013})  
	\begin{eqnarray}\label{equ2}
		\mathcal{I}_{k}(\infty)=\frac{\pi^{k}}{{2}^{2k-1}k!}.
	\end{eqnarray}  
	A direct proof was very recently provided\cite{Dominik2025}. The aforementioned works exclusively focus on infinite-time solutions. Till date no finite-time solution has been reported even by using Physics arguments as in Refs.\onlinecite{Kholodenko2012, Kenmoe2013}. This is likely due to the fact that the finite-time solution to the LZ problem involves higher transcendental special functions\cite{Abramowitz,Wong2000, Fai2024}--namely, Parabolic Cylinder Weber functions (PCFs), Hypergeometric functions or Hermite's polynomials whereas the infinite-time involves a simple exponential of the so called LZ parameter. 
	
	In this work, by further leveraging the deep connection with the LZ problem, we propose a closed-form finite-time solution to Eq.\eqref{equ1} and simultaneously derive various integral representations for the product of PCFs. In order to validate our approach, our analytical solutions are compared with data from numerical simulations of the original integrals. Both data are found to be in excellent quantitative agreement confirming the robustness of our analytical approach. Our findings  offer new opportunities to gain further insight into the non-trivial mechanism behind level-crossing phenomena, particularly those that remain incompletely understood such as the SU(3) LZ problem\cite{Kiselev2013},  classical slow noise-induced finite-time LZ transitions\cite{Kenmoe2013, Nyisomeh2019, Luo2017}, and Many-body LZ effects\cite{Garanin2005}. Additionally, real systems unavoidably evolve over bounded time intervals where finite time-evolution operators can radically differ from their long-time asymptotic limits. In experiments involving for example, qubits and/or qutrits, transitions are often stopped or measured at finite times.  The finite-time solution $\mathcal{I}_{k}(\tau)$ is therefore of paramount importance not just in level-crossing problems but clearly beyond. This includes cosmology\cite{Enomoto2022} and production of topological defects\cite{Damski2005, Kou2025}, and quantum optics\cite{Saito2006}. This approach can be extended to more complex time-dependent systems allowing solving more complicated integrals.
	
	The rest of the paper is organized as follows. In Section \ref{Sec2} the main results are presented while detailed proofs are provided in \ref{Sec3}. Section \ref{Sec4} illustrates the specific cases of $\mathcal{I}_{4}(\tau)$ and $\mathcal{I}_{5}(\tau)$ which are numerically verified using {\it Mathematica}. A conclusion drawing our achievements can be found in Section \ref{Sec6}.

	\section{Main Results}\label{Sec2}
	
	At finite times, based on available techniques in both Mathematical analysis and theoretical Physics, one can only calculate $\mathcal{I}_{1}(\tau)$. Indeed, $\mathcal{I}_{1}(\tau)$ is widely known in Physics in the context of level-crossing phenomena. It can be calculated by opening up the unique cosine function and realizing that each of the resulting integrals is symmetric by interchanging the arguments of the cosines. This allows to reduce the double integral to a square of a single integral in the sense of Dyson series. The result generally involves Fresnel integrals\cite{Kenmoe2013, Kiselev2013} (see Section \ref{Sec4.2}). However, the method elaborated in this work expresses $\mathcal{I}_{1}(\tau)$ in terms of Parabolic Cylinder Functions (PCFs) as
	\begin{eqnarray}\label{equ3}
		\mathcal{I}_{1}(\tau)=\int^\tau_{-\infty}d\tau_{1}\int^{\tau_{1}}_{-\infty}d\tau_{2}\cos[\tau^{2}_{1}-\tau^{2}_{2}]=\frac{1}{4}{\left|D_{-1}\left(-i\mu_0\tau\right)\right|}^{2},
	\end{eqnarray}  
	leading to a connection between Fresnel integrals and PCFs (see Eq.\eqref{equ4.15}). Here, $\mu_0=\sqrt{2}e^{-i\pi/4}$ and $D_{\nu}(z)$ the PCF of order $\nu$ and argument $z$ with the special case $D_{-1}(z)=e^{z^2/4}\sqrt{\frac{\pi}{2}}[1-{\rm erf}(z/\sqrt{2})]$ where ${\rm erf}(z)$ stands for the error function\cite{Abramowitz, Fai2024}. It possesses the integral representation,
	$D_{-1}\left(-i\mu_0\tau\right)=i\mu_0e^{i\tau^2/2}\int^\tau_{-\infty}d\tau_{1}e^{-i\tau^2_1}.$
	At large negative times $\tau\to-\infty$ the function obviously collapses i.e. $D_{-1}\left(-i\mu_0\tau\right)\sim 0$ while at large positive times $\tau\to\infty$, it asymptotically approaches $D_{-1}\left(-i\mu_0\tau\right)\sim e^{-i\tau^2/2}\sqrt{2\pi}+\mathcal{O}\left(\tau^{-1}\right)$ and $\left|D_{-1}\left(-i\mu_0\tau\right)\right|\sim \sqrt{2\pi}$ yielding the result in \eqref{equ2}. At the level crossing $\tau=0$, it acquires the exact value $D_{-1}\left(0\right)=\sqrt{\pi/2}$ and $\mathcal{I}_{1}(0)=\pi/8$ (see Eq.\eqref{equ14b}). $\mathcal{I}_{1}(\tau)$ depicts the gross temporal profile of LZ transition probabilities (see Figure \ref{Figure0}$(d)$). Thus, with the above values,  the transition time\cite{Vitanov1999}: $\tau_{\rm tr}=\mathcal{I}_{1}(\infty)/\mathcal{I}'_{1}(0)$ can be calculated (with the prime denoting time derivative). A simple calculation shows that $\tau_{\rm tr}=1$.
	
	The sine version of  \eqref{equ3} denoted as $\mathcal{J}_{1}(\tau)$ is presented in Eq.\eqref{equ4.12c}. It is also desired in level-crossing Physics. It vanishes at $\tau=\infty$ as a result of the integration of an odd function between symmetric boundaries. The finite time case, is calculated in Section \ref{Sec4.2}. Therein, $\mathcal{I}_{1}(\tau)$ is represented in terms of a Hypergeometric function\cite{Abramowitz, Fai2024}. The trajectory of $\mathcal{Z}(\tau)=\mathcal{I}_{1}(\tau)+i\mathcal{J}_{1}(\tau)$ is depicted in Figure.\ref{Figure0}. It describes a classical particle rolling spirally on the internal surface of a cylinder.
	
	\begin{figure}[]
		\vspace{-0.5cm}
		\centering
		\begin{center} 
			\includegraphics[width=12cm, height=7cm]{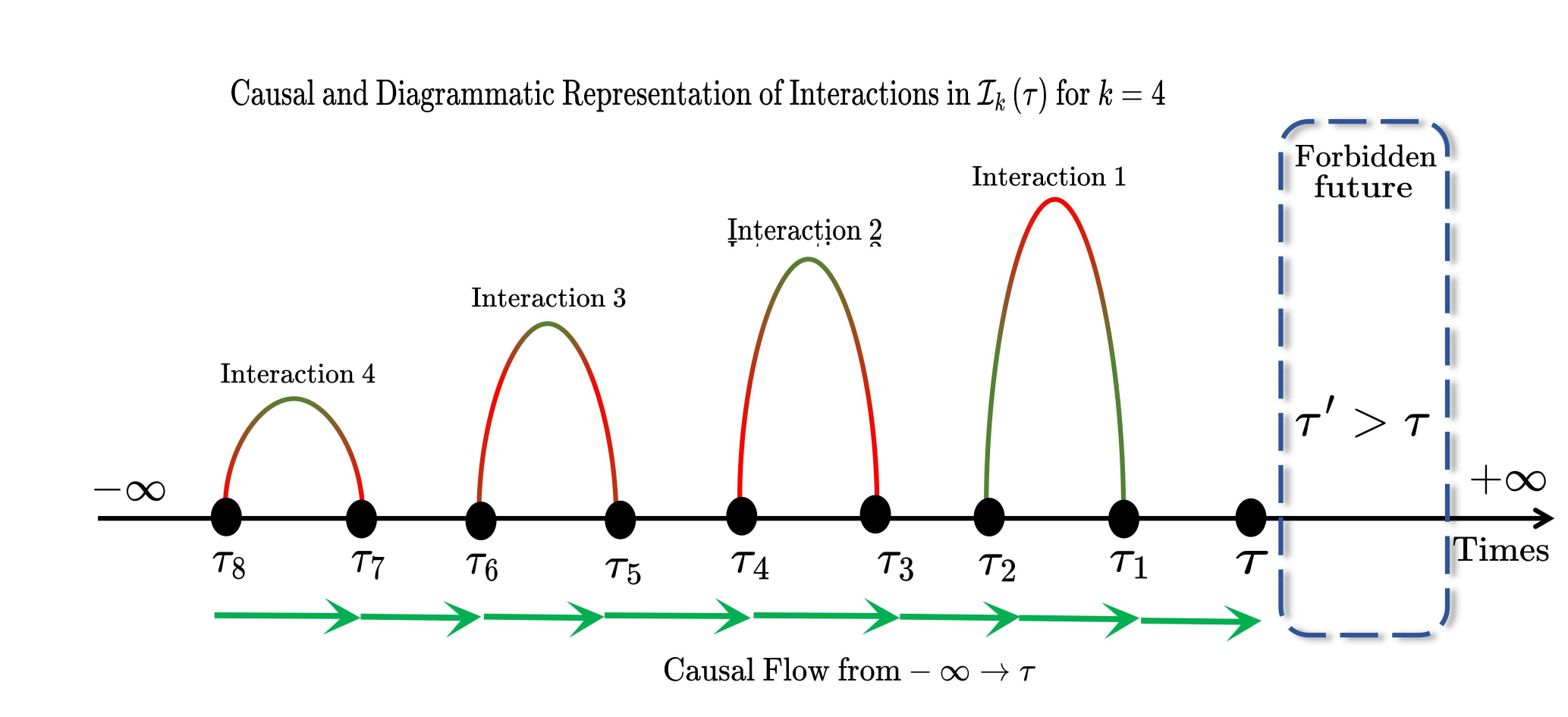}
			\vspace{-0.5cm}
			\caption{Diagrammatic representation of the fourth order integral in the time-domain $-\infty\leq\tau_8\leq\tau_7\leq\cdots\leq\tau_2\leq\tau_1\leq\tau$. Black dots denote $\tau_1$, $\tau_2$, ...,$\tau_8$. Red arcs describe interactions between the paths $\left(\tau_{2j-1},\tau_{2j}\right)$ and represent cosines kernels. Green arrows encode the constraints $\tau_{2j}<\tau_{2j-1}$. The architecture generated by arrows describe time-ordered causal structure with four virtual interactions.  The structure of the integral is time non-local i.e. the value at later time $\tau$ depends on the whole history starting from $\tau=-\infty$ through intermediate times. It is also causal (nothing travels back in time) i.e. it does not depends on the future time $\tau'>\tau$. These characterize systems with temporal memory.}
			\label{Figure00}
		\end{center}
	\end{figure}
	
	In level-crossing Physics, the integrand $\cos[\tau^{2}_{1}-\tau^{2}_{2}]$ arises from time-dependent transition amplitudes. The  zero-coupling energies $E_{\pm}=\pm \tau$ are trajectories (time paths) that cross at $\tau=0$ (level crossing) in the energy diagram $\left(\tau,E\right)$. As a system evolves along these paths, the quantity $\tau^{2}=2\int^{\tau}\tau d\tau$ represents the dynamical phase (semi-classical action) accumulated  by the system when traversing the level-crossing up to a time $\tau$. The phase difference $\tau^{2}_{1}-\tau^{2}_{2}$ therefore stands for the action difference between two paths evolving from different instant $\tau_2$ and $\tau_1$. The cosine then reflects the interactions between these paths. In general, we can assert that $\cos[\tau^{2}_{1}-\tau^{2}_{2}]$ encapsulates {\it one virtual} interaction between two intersecting paths. This mechanism is illustrated in Figure \ref{Figure00}. 
	
	At a deeper level, the level crossing acts as a beam-splitter: an incoming wave can partially be transmitted or partially reflected. The integral $\mathcal{I}_{1}(\tau)$ can then be viewed as a wave that traverses the beam-splitter. Since there is only one virtual interaction and no recombination occurs, no interference pattern can emerge. In contrast, higher-order such as $\mathcal{I}_{2}(\tau)$ which involve two cosines kernels, represents two virtual interactions. A wave can split at one level-crossing and recombines at another giving rise to  quantum interferences that can be constructive or destructive.
	
	In the American Mathematical Monthly 2012 challenge\cite{Silagadze2012}, Silagadze asked to evaluate $\mathcal{I}_{2}(\infty)$. This was done and its generalization $\mathcal{I}_{k}(\infty)$ to arbitrary $k$ is today known in Mathematics as the Kholodenko-Silagadze integral\cite{Kholodenko2012, Dominik2025, Juan2015, Richard2014}. The finite-time indirect calculation based on Physics arguments allowed us to come up with
	\begin{eqnarray}\label{equ4}
		\nonumber\mathcal{I}_{2}(\tau)&=&\int^\tau_{-\infty}d\tau_{1}\int^{\tau_{1}}_{-\infty}d\tau_{2}\int^{\tau_{2}}_{-\infty}d\tau_{3}\int^{\tau_{3}}_{-\infty}d\tau_{4}\cos[\tau^{2}_{1}-\tau^{2}_{2}]\cos[\tau^{2}_{3}-\tau^{2}_{4}],\\
		&=&\frac{1}{64}\left(\pi\left|D_{-1}\left(-i\mu_{0}\tau\right)\right|^{2}+4\mathrm{Im}\left[D_{-1}\left(-i\mu_{0}\tau\right)\mathcal{D}^{\left(1\right)}_{-1}\left(i\overline\mu_0\tau\right)\right]\right). 
	\end{eqnarray}   
	Here, $\bar{\mu}_0=\sqrt{2}e^{i\pi/4}$ and  $\mathcal{D}^{\left(1\right)}_{\nu}(z)$ is the second modified PCF of order $\nu$ and argument $z$. As usual, $\mathrm{Im}(\cdots)$ denotes imaginary part of functions. The generalized function $\mathcal{D}^{\left(r\right)}_{\nu}(z)$  including $\mathcal{D}^{\left(1\right)}_{\nu}(z)$ for all non-negative integer $r$ are presented in Section \ref{Sec3}. The integrand $\cos[\tau^{2}_{1}-\tau^{2}_{2}]\cos[\tau^{2}_{3}-\tau^{2}_{4}]$ encodes two virtual interactions between two independent pair of paths $(\tau_1,\tau_2)$ and $(\tau_3,\tau_4)$. The processes evolve in parallel. While the mathematical structures separate the paths into two pairs, the integration over the nested time domain causes coupling between them leading to quantum interference. $\mathcal{I}_{2}(\tau)$ then describes the superposition of two waves,  $\mathcal{I}^{\left(1\right)}_{2}(\tau)=\pi\left|D_{-1}\left(-i\mu_{0}\tau\right)\right|^{2}/64$ and $\mathcal{I}^{\left(2\right)}_{2}(\tau)=\mathrm{Im}\left[D_{-1}\left(-i\mu_{0}\tau\right)\mathcal{D}^{\left(1\right)}_{-1}\left(i\overline\mu_0\tau\right)\right]/16$ (see Insert in Figure.\ref{Figure2}$(a)$). The wave $\mathcal{I}^{\left(1\right)}_{2}(\tau)$ passes through the first level-crossing and meets $\mathcal{I}^{\left(2\right)}_{2}(\tau)$ at the second one. Because $\mathcal{I}_{2}(\tau)>\mathcal{I}^{(2)}_{2}(\tau)>\mathcal{I}^{(1)}_{2}(\tau)$ and $\mathcal{I}^{(2)}_{2}(\tau)+\mathcal{I}^{(1)}_{2}(\tau)=\mathcal{I}_{2}(\tau)$, 
	the two waves interfere constructively. Transitions then arise by constructive interference between virtual paths.
	
	Our approach consists of mapping $\mathcal{I}_{k}\left(\tau\right)$ onto the population difference (density matrix) in the two-level LZ problem and by further Taylor expanding  the PCFs with respect to the index about $\nu=0$  and finally comparing the resulting series. 
	After paying the cost of this efforts, we arrived at
	\begin{eqnarray}\label{equ5}
		\mathcal{I}_{k}\left(\tau\right)=\frac{\pi^{k-1}}{2^{4k-2}}\sum^{k}_{n=0}\frac{\left(-\frac{2}{\pi}\right)^n }{n!\left(k-n-1\right)!}\mathcal{J}_n\left(\tau\right),\quad k>0.
	\end{eqnarray}
	This formula is our main result. Here, 
	\begin{eqnarray}\label{equ6}
		\mathcal{J}_n\left(\tau\right)=\sum_{j=0}^{n}\sum_{k=0}^{n-j}\sum_{\ell=0}^{j}\binom{n}{j}\mathcal{P}_{n-j,k}\left(0\right)\mathcal{P}^{*}_{j,\ell}\left(0\right)\mathcal{D}^{\left(k\right)}_{-1}\left(-i\mu_0\tau\right)\mathcal{D}^{\left(\ell\right)}_{-1}\left(i\bar{\mu}_0\tau\right),
	\end{eqnarray} 
	where 
	\begin{eqnarray}\label{equ7}
		\mathcal{P}_{n,m}\left(\nu\right) = (-i)^{m}\binom{n}{m}\sum_{j=1}^{n-m}\mathcal{B}_{n-m,j}\left(\phi^{\left(1\right)},\phi^{\left(2\right)},\cdots,\phi^{\left(n-m-1\right)},\phi^{\left(n-m\right)}\right),
	\end{eqnarray} 
	are polynomials of $\phi^{\left(r\right)}$ with $\phi^{\left(1\right)}\left(i\nu+1\right)\equiv -i\left(\psi^{\left(0\right)}\left(i\nu+1\right)-\frac{\ln 2}{2}\right)$ and $\phi^{\left(r\right)}\left(i\nu+1\right)\equiv-i^{r}\psi^{\left(r-1\right)}\left(i\nu+1\right)$ for $r>1$ where $\psi^{\left(r\right)}$ are polygamma functions\cite{Abramowitz, Wong2000, Fai2024} and $\mathcal{B}_{n,m}$ the partial Bell polynomials\cite{comtet1974}. A few exceptions to the polynomials in Eq.\eqref{equ7} are in order. $\mathcal{P}_{n,m}\left(\nu\right)$ vanish for all pair $(n,m)$ of non-negative integers for which $ n<m$ or $m<0$ and $\mathcal{P}_{n,n}\left(\nu\right)=(-i)^n$ for arbitrary $\nu$ including $\nu=0$.
	
	The first few order of $\mathcal{I}_{k}\left(\tau\right)$ as presented in \eqref{equ5} can easily be manually inferred. Higher order can also be calculated from \eqref{equ5} by using symbolic calculators such as Mathematica or Maple. Based on  \eqref{equ5}, a dedicated Mathematica package named  {\bf OscillatoryIntegralAnalytical.wl} for automatically generating $\mathcal{I}_{k}(\tau)$ symbolically for all $k>0$ is developed and placed in our \href{https://github.com/Kenmax15/Closed-Form-Approach-to-Oscillatory-Integrals/tree/main}{GitHub repository}. The package is publicly accessible under a GNU general public licence. It is simple, easy to use (see README file) and works on Mathematica $\ge $ 10 and perhaps earlier versions. For $k=3$  for example, the package returns,
	\begin{eqnarray}\label{equ7a}
		\nonumber\mathcal{I}_{3}(\tau)&=&\int^\tau_{-\infty}d\tau_{1}\int^{\tau_{1}}_{-\infty}d\tau_{2}\int^{\tau_{2}}_{-\infty}d\tau_{3}\int^{\tau_{3}}_{-\infty}d\tau_{4}\int^{\tau_{4}}_{-\infty}d\tau_{5}\int^{\tau_{5}}_{-\infty}d\tau_{6}\cos[\tau^{2}_{1}-\tau^{2}_{2}]\cos[\tau^{2}_{3}-\tau^{2}_{4}]\cos[\tau^{2}_{5}-\tau^{2}_{6}]\\
		\nonumber	&=&\frac{1}{2^8.3!}\Big[\frac{7\pi^2}{4}\left|D_{-1}\left(-i\mu_{0}\tau\right)\right|^{2}+6\left|\mathcal{D}^{\left(1\right)}_{-1}\left(-i\mu_{0}\tau\right)\right|^{2}+6\pi\mathrm{Im}\left[D_{-1}\left(-i\mu_{0}\tau\right)\mathcal{D}^{\left(1\right)}_{-1}\left(i\overline\mu_0\tau\right)\right]\\&-&6\mathrm{Re}\left[D_{-1}\left(-i\mu_{0}\tau\right)\mathcal{D}^{\left(2\right)}_{-1}\left(i\overline\mu_0\tau\right)\right]\Big],
	\end{eqnarray} 
	which corresponds to the result obtained by manual derivations. ${\rm Re}(\cdots)$ designates real parts of functions.
	Our results are all verified both analytically and numerically. In Figure.\ref{Figure2}, a benchmarking of the   analytical and numerical results is presented. We observe a perfect qualitative and quantitative agreement between both data confirming our analytical method.

	\begin{figure}[]
		\vspace{-0.5cm}
		\centering
		\begin{center} 
			\includegraphics[width=7.8cm, height=6cm]{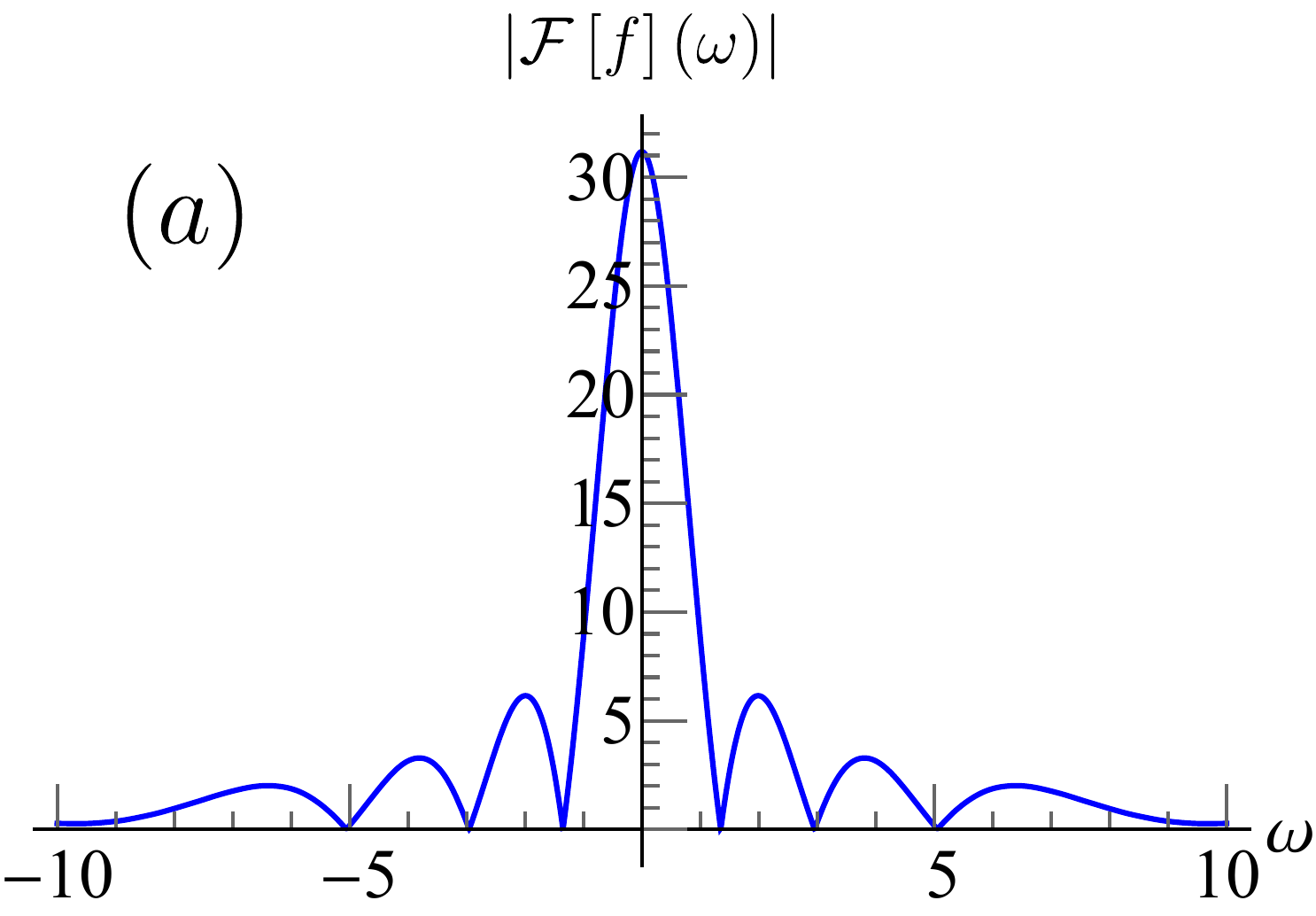}\hspace{-0.75cm}
			\includegraphics[width=7.8cm, height=6cm]{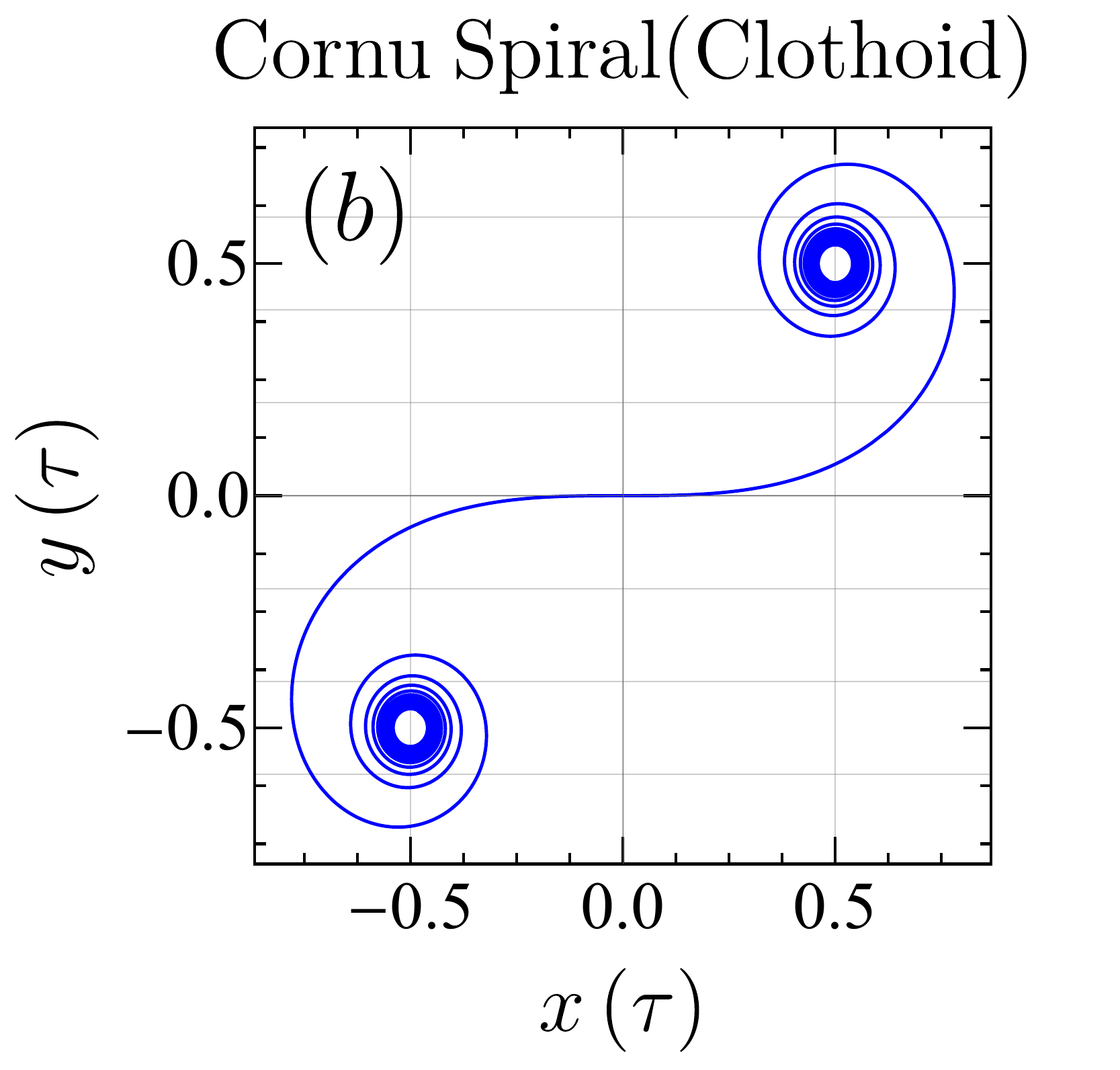}\\ 
			\includegraphics[width=7.8cm, height=6cm]{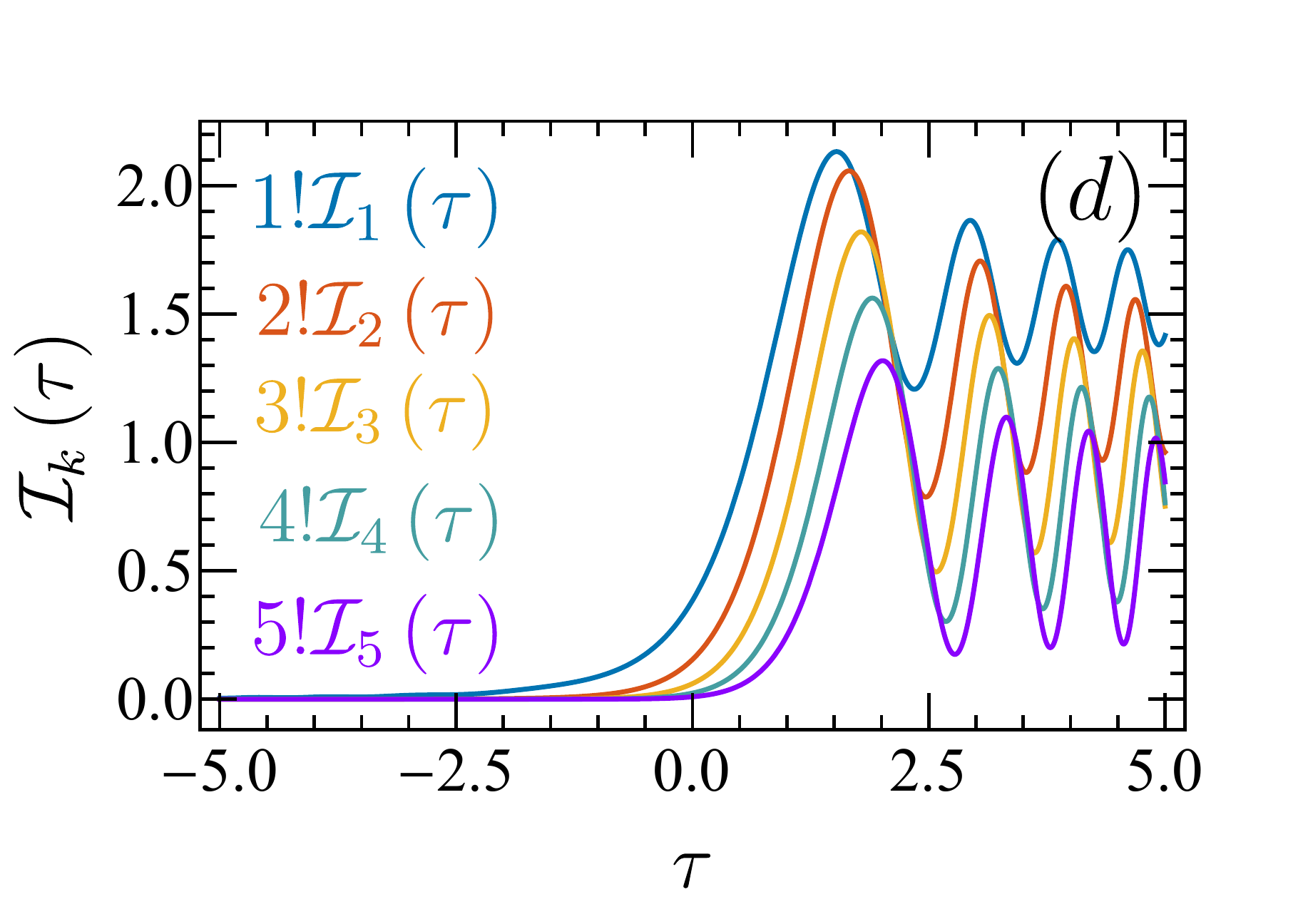}\hspace{-0.75cm}
			\includegraphics[width=7.8cm, height=6cm]{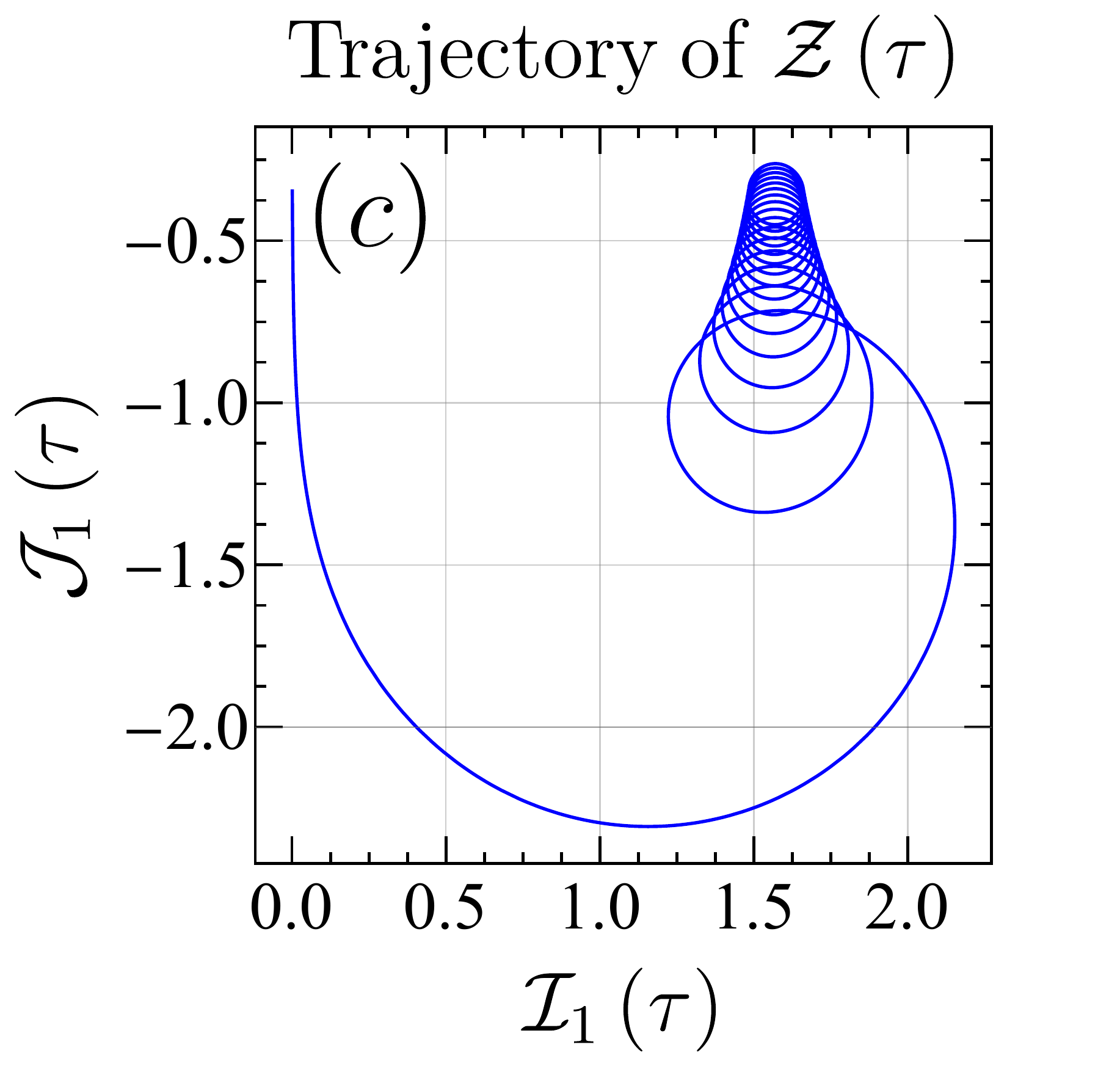}
			\vspace{-0.3cm}
			\caption{Visualizations of $\mathcal{I}_{1}(\tau)$ and/or $\mathcal{J}_{1}(\tau)$. $(a)$ Modulus of the Fourier transform  $\mathcal{F}[f](\omega)=\int_{-\infty}^{\infty}f(\tau)e^{i\omega \tau}d\tau$ with $f(\tau)=\left|D_{-1}\left(-i\mu_0\tau\right)\right|^{2}$ where $\omega$ is the frequency. This is a visualization of $\mathcal{I}_{1}(\tau)$ from the frequency-domain. The resulting spectrum resembles a sinus cardinal function with multiple zeros along the $\omega$-axis. $(b)$ Parametric plot of   $\left|D_{-1}\left(-i\mu_0\tau\right)\right|^{2}=\pi({[\frac{1}{2}+C(\sqrt{2/\pi}\tau)]}^{2}+{[\frac{1}{2}+S(\sqrt{2/\pi}\tau)]}^{2})$. 
				For calculations, we have defined $x\left(\tau\right)=C(\sqrt{2/\pi}\tau)$ and $y\left(\tau\right)=S(\sqrt{2/\pi}\tau)$. Thus, $r^2(\tau)=\left(x(\tau)+\frac{1}{2}\right)^2+\left(y(\tau)+\frac{1}{2}\right)^2$ is the square of the Euclidean distance between the points $(x\left(\tau\right), y\left(\tau\right))$ and $(-1/2, -1/2)$. Therefore, defining again $\mathcal{A}(\tau)=\left|D_{-1}\left(-i\mu_0\tau\right)\right|^{2}$ the above equation takes the form $\mathcal{A}(\tau)=\pi r^2(\tau)$ telling us that geometrically, $\mathcal{I}_{1}(\tau)$ can be regarded as the area of a parametric  circle of varying radius centered at the point $(-1/2,-1/2)$. $(c)$ Trajectory of $\mathcal{Z}(\tau)=\mathcal{I}_{1}(\tau)+i\mathcal{J}_{1}(\tau)$ (see also Ref.\onlinecite{Glasbrenner2023}). We see that it describes the motion of a classical particle spiraling on the inner surface of a cylinder. The integral $\mathcal{J}_{1}(\tau)$ is presented in Eq.\eqref{equ4.12c}. (d) Scaled integrals $k!\mathcal{I}_{k}(\tau)$ for $k=\{1,2,3,4,5\}$}
			\label{Figure0}
		\end{center}
	\end{figure}
	
	\section{Proof}\label{Sec3}
	As the problem of central interest in this note can not be directly solved using actual mathematical techniques and given its intimate connection with an existing exactly solvable non-stationary quantum mechanical problem, we exploit this interplay and obtained a closed-form analytical solution. In this Section, we present the relevant details of our derivations. A numerical analysis supporting our results is also presented.

	\subsection{Tools}
	The workhorse used to evaluate $\mathcal{I}_{k}\left(\tau\right)$ is the LZ theory\cite{Landau1932, Zener1932, Stuckelberg1932, Majorana1932}. For the sake of clarity, it is essential to briefly review this theory. Consider two diabatics (uncoupled) states $\ket{1}$ and $\ket{2}$  driven in the quantization direction by a linear magnetic field that changes its sign at a level-crossing at a non-negative sweep rate $\alpha>0$. Suppose the dynamics are governed by the time-dependent Schr\"odinger equation for the probability amplitudes $C_1(\tau)$ and $C_2(\tau)$ respectively. In this situation, the system dwells in its initial state in the course of time. Now, let us couple the two states such that the inter-level distance between them is $\Delta$. The diabatic states hybridize at level-crossing leading to the time-dependent adiabatic states $\ket{\psi_1(\tau)}$ and $\ket{\psi_2(\tau)}$ that come close at $\tau=0$ forming an avoided level-crossing.  These states operate as shuttle mediating population transfer between diabatic states. This is the LZ phenomenon.  Finding the probabilities to remain in the same diabatic state or to transition to a different one is the LZ problem\cite{Landau1932, Zener1932, Stuckelberg1932, Majorana1932}. It can be described from the optical Bloch picture by setting a Bloch vector $\vec{u}(\tau)=\left[u_x(\tau), u_y(\tau),u_{z}(\tau)\right]^\mathcal{T}$ on the surface of a unit-radius  sphere where $u_x(\tau)=2{\rm Re}(\boldsymbol{\rho}_{12}(\tau))$, $u_y(\tau)=2{\rm Im}(\boldsymbol{\rho}_{12}(\tau))$, and $u_z(\tau)=\rho_{11}(\tau)-\rho_{22}(\tau)$ with $\mathcal{T}$ denoting transposed vector and $\boldsymbol{\rho}_{ij}(\tau)$ the matrix elements of the density matrix $\boldsymbol{\rho}(\tau)$ with indices $i,j=(1,2)$ standing for diabatic states. 
	Here, $\rho_{11}(\tau)=|C_1(\tau)|^2$ and $\rho_{22}(\tau)=|C_2(\tau)|^2$ are level populations and $\boldsymbol{\rho}_{12}(\tau)=C_1(\tau)C^{*}_2(\tau)$ the coherence factor with $\boldsymbol{\rho}_{21}(\tau)=\boldsymbol{\rho}^{*}_{12}(\tau)$ where the star denotes the complex conjugate; 
	\begin{eqnarray}\label{equ8aa}
		C_1(\tau)=-e^{-\pi\nu/4}e^{-i\pi/4} e^{i\varphi_1}D_{-i\nu}\left(-i\mu_0\tau\right),
	\end{eqnarray} 
	and  
	\begin{eqnarray}\label{equ8ab}
		C_2(\tau)=\sqrt{\nu}e^{-\pi\nu/4} e^{i\varphi_2}D_{-i\nu-1}\left(-i\mu_0\tau\right),
	\end{eqnarray} 
	are respectively the probability amplitudes to remain in the state $\ket{1}$ and to transition to $\ket{2}$ at time $\tau$ provided that the system is prepared at a far remote negative time in the state $\ket{1}$;  $\nu=\Delta^2/2\alpha$ is the LZ parameter\cite{Kenmoe2013, Vitanov1996}. The angles $\varphi_{1,2}$ are arbitrary phases that play no role in probabilities. They are however inserted in order to ensure that the real and imaginary parts of the complex numbers $C_{1,2}(\tau)$ match numerical solutions.
	
	The temporal evolution of the components of the Bloch vector read $\dot{u}_x(\tau)=-2\tau u_y(\tau)$, $\dot{u}_y(\tau)=2\tau u_x(\tau)-2\sqrt{2\nu}u_z(\tau)$, and $u_z(\tau)=2\sqrt{2\nu}u_y(\tau)$. Hereby, with initial conditions $u_x(-\infty)=u_y(-\infty)=0$ and $u_z(-\infty)=1$, these equations can be re-written in integral-equation forms as\cite{Kenmoe2013, Vitanov1996, Kou2025}
	\begin{eqnarray}\label{equ8a}
		u_x\left(\tau\right)=2\sqrt{2\nu}\int^{\tau}_{-\infty}d\tau_1\sin[\tau^{2}-\tau^{2}_{1}]u_z(\tau_{1}),
	\end{eqnarray}  
	\begin{eqnarray}\label{equ8b}
		u_y\left(\tau\right)=-2\sqrt{2\nu}\int^{\tau}_{-\infty}d\tau_1\cos[\tau^{2}-\tau^{2}_{1}]u_z(\tau_{1}),
	\end{eqnarray} 
	and 
	\begin{eqnarray}\label{equ8}
		u_z\left(\tau\right)=1-\left(2\sqrt{2\nu}\right)^{2}\int^{\tau}_{-\infty}d\tau_1\int_{-\infty}^{\tau_1}d\tau_2\cos[\tau_1^{2}-\tau^{2}_{2}]u_z(\tau_{2}).
	\end{eqnarray}  
	The $\tau^{2}$-dependence results from the dynamical phase accumulated by the system when traversing the level-crossing. These integral equations are difficult to evaluate  directly both analytically and numerically. In the later case, the kernel highly oscillates, numerical integrations are slow and unstable giving low-precision solutions. One strategy in avoiding this issue consists of reducing them to third order linear ODEs (see Refs.\onlinecite{Kenmoe2013, Vitanov1996} and Section \ref{Sec2b}). Numerically, ODEs are more stable than their integral counterparts.  Analytically, the resulting ODE are indirectly solved by invoking the fact that $u_z(\tau)$ evolves within the SO(3) group while the probability amplitudes $C_{1,2}(\tau)$ evolve within SU(2). Thus, because SU(2) doubly covers SO(3), every element in SO(3) correspond to two elements in SU(2), meaning that the dynamics in SU(2) can fully capture the behavior in SO(3).  For the purpose of this work, Eq.\eqref{equ8} is solved iteratively\cite{Kenmoe2013},
	\begin{eqnarray}\label{equ9}
		u_z\left(\tau\right)=\sum^{\infty}_{k=0}{{\left(-\nu\right)}^{k}}{2}^{3k}\mathcal{I}_{k}\left(\tau\right).
	\end{eqnarray}
	The desired integrals immediately appear here. $\mathcal{I}_k\left(\tau\right)$ is the $k$th contribution to the infinite-order perturbation series for the population difference between two crossing levels.  The LZ parameter is our perturbation parameter. For instance, $\mathcal{I}_{1}\left(\tau\right)$ as the first order contribution in a perturbation theory, tells us how $u_z\left(\tau\right)$ starts to grow when the system is slightly perturbed from the diabatic evolution (zero coupling, $\nu=0$). Therefore, knowing $u_z(\tau)$ is sufficient to determine $\mathcal{I}_{k}\left(\tau\right)$ provided its Taylor series expansion with respect to $\nu$ about $\nu=0$ (Maclaurin series) can be calculated. This is one of the main challenges we have to tackle. Indeed, in the infinite-time limit $u_z\left(\infty\right)=2e^{-\pi\nu}-1$ and a Taylor's expansion is easy. At finite times, however, this task becomes tricky since
	\begin{eqnarray}\label{equ10}
		u_z(\tau)=-\nu e^{-\pi\nu/2}\left[\left|D_{-i\nu-1}\left(-i\mu_0\tau\right)\right|^{2}-\frac{1}{\nu}\left|D_{-i\nu}(-i\mu_0\tau)\right|^{2}\right]=1-2P_{\rm LZ}(\tau,\nu),
	\end{eqnarray} 
	where
	\begin{eqnarray}\label{equ11}
		P_{\rm LZ}(\tau,\nu)=\nu e^{-\pi\nu/2}\left|D_{-i\nu-1}\left(-i\sqrt{2}e^{-i\pi/4}\tau\right)\right|^{2},
	\end{eqnarray} 
	is nothing but the finite-time LZ transition probability\cite{Kenmoe2013, Vitanov1996}. The last equality in \eqref{equ10} is obtained by using \eqref{equ4.17}. Therein, $P_{\rm LZ}(\tau,\nu)$ corresponds to the probability for finding the system in the state $\ket{2}$ at a finite time $\tau$ provided it stated off at $\tau=-\infty$ in the state  $\ket{1}$. It is  an infinitely differentiable smooth function of the real variable $\nu$ at $\nu=0$ and for this reason admits a one-dimensional Taylor (Maclaurin) series expansion,
	\begin{eqnarray}\label{equ12}
		P_{\rm LZ}(\tau,\nu)=\sum^{\infty}_{k=0}\frac{1}{k!}\left(\frac{\partial^{k}P_{\rm LZ}(\tau,\nu)}{\partial \nu^{k}}\Big|_{\nu=0}\right)\nu^{k},
	\end{eqnarray}
	the radius of convergence of which is infinite and the series is absolutely convergent. Such a representation is highly desired in the Physics of noise-induced LZ transitions as discussed in Refs.\onlinecite{Kenmoe2013, Luo2017}. Indeed, the classical slow noise-induced LZ transition probability is obtained by averaging $P_{\rm LZ}(\tau,\nu)$ over the ensemble distribution of the noise by replacing in the LZ parameter, the inter-level distance by a noise field $Q$ as $\nu=Q^2/2\alpha$. This is not possible with the finite-time solution \eqref{equ11} as $\nu$ enters it in a non-trivial way. Moreover, in the representation \eqref{equ12} this becomes feasible. As yet another consequence stemming from \eqref{equ12} a Fourier transform of $P_{\rm LZ}(\tau,\nu)$ with respect to $\nu$ may be performed allowing to measure how sharply or smoothly the system responds to the change in couplings.
	
	We now have to mention that the derivatives in \eqref{equ12} are difficult to calculate mainly because PCFs do not have simple derivatives with respect to the index. Assuming that this can be circumvent, it might have several physical consequences in deciphering observables that can be expressed in terms of $\left|D_{-i\nu-1}\left(-i\mu_0\tau\right)\right|^2$ . For instance, the extremal values of $P_{\rm LZ}(\tau,\nu)$ can be determined  by solving the equation $\partial P_{\rm LZ}(\tau,\nu)\partial\nu=0$ for $\nu$ while inflection points (points where the function changes its concavity) are measured by analyzing $\partial^2 P_{\rm LZ}(\tau,\nu)/\partial\nu^2$ at the vicinity of $\nu_{\rm infl}$. When $\partial^2 P_{\rm LZ}(\tau,\nu)/\partial\nu^2>0$ the probability is concave up and concave down in the opposite case.  Thus, after calculating all derivatives including higher order, then substituting \eqref{equ12} into \eqref{equ9} and comparing the resulting two series we obtain,
	\begin{eqnarray}\label{equ13}
		\mathcal{I}_{k}\left(\tau\right)=\frac{\left(-1\right)^{k+1}}{2^{3k-1}k!}\left(\frac{\partial^{k}P_{\rm LZ}(\tau,\nu)}{\partial \nu^{k}}\Big|_{\nu=0}\right).
	\end{eqnarray}
	We observe that the integrals $\mathcal{I}_{k}\left(\tau\right)$ can also be interpreted as the rates of change of $P_{\rm LZ}(\tau,\nu)$ and its derivatives at $\nu=0$ in the course of time. For instance, $\mathcal{I}_{1}\left(\tau\right)$ and $\mathcal{I}_{2}\left(\tau\right)$ are up to some constants the rates of change of $P_{\rm LZ}(\tau,\nu)$ and $\partial P_{\rm LZ}(\tau,\nu)/\partial\nu$ at $\nu=0$ respectively.   Geometrically, $\mathcal{I}_{1}\left(\tau\right)$ it is the tangent to the curve of $P_{\rm LZ}(\tau,\nu)$ at $\nu=0$.  
	
	Achieving finite time solutions as claimed so far depends on our ability to calculate the right-hand side of \eqref{equ13}. Thus, using the Leibniz formula for higher order derivatives of the product of two functions, one finds that
	\begin{eqnarray}\label{equ15}
		\frac{\partial^{k}P_{\rm LZ}(\tau,\nu)}{\partial \nu^{k}}=e^{-\pi\nu/2}\sum_{n=0}^{k}\binom{k}{n}
		\left(-\frac{\pi}{2}\right)^{k-n}\left(\nu-\frac{2(k-n)}{\pi}\right)\frac{\partial^{n}\Big|D_{-i\nu-1}(-i\mu_{0}\tau)\Big|^2}{\partial\nu^{n}}.
	\end{eqnarray}
	From here, our task reduces to computing the higher order derivatives of $D_{-i\nu-1}(-i\mu_{0}\tau)$ with respect to $\nu$. A formula yielding this for arbitrary $\nu$ can be found in Ref.\onlinecite{Mathematica}. Unfortunately, this formula is extraordinary complex involving complicated special functions that are cumbersome and difficult to manipulate.  A simpler formula is derived in the coming Section.
	
	\begin{figure}[]
		\vspace{-0.5cm}
		\centering
		\begin{center} 
			\includegraphics[width=8cm, height=5cm]{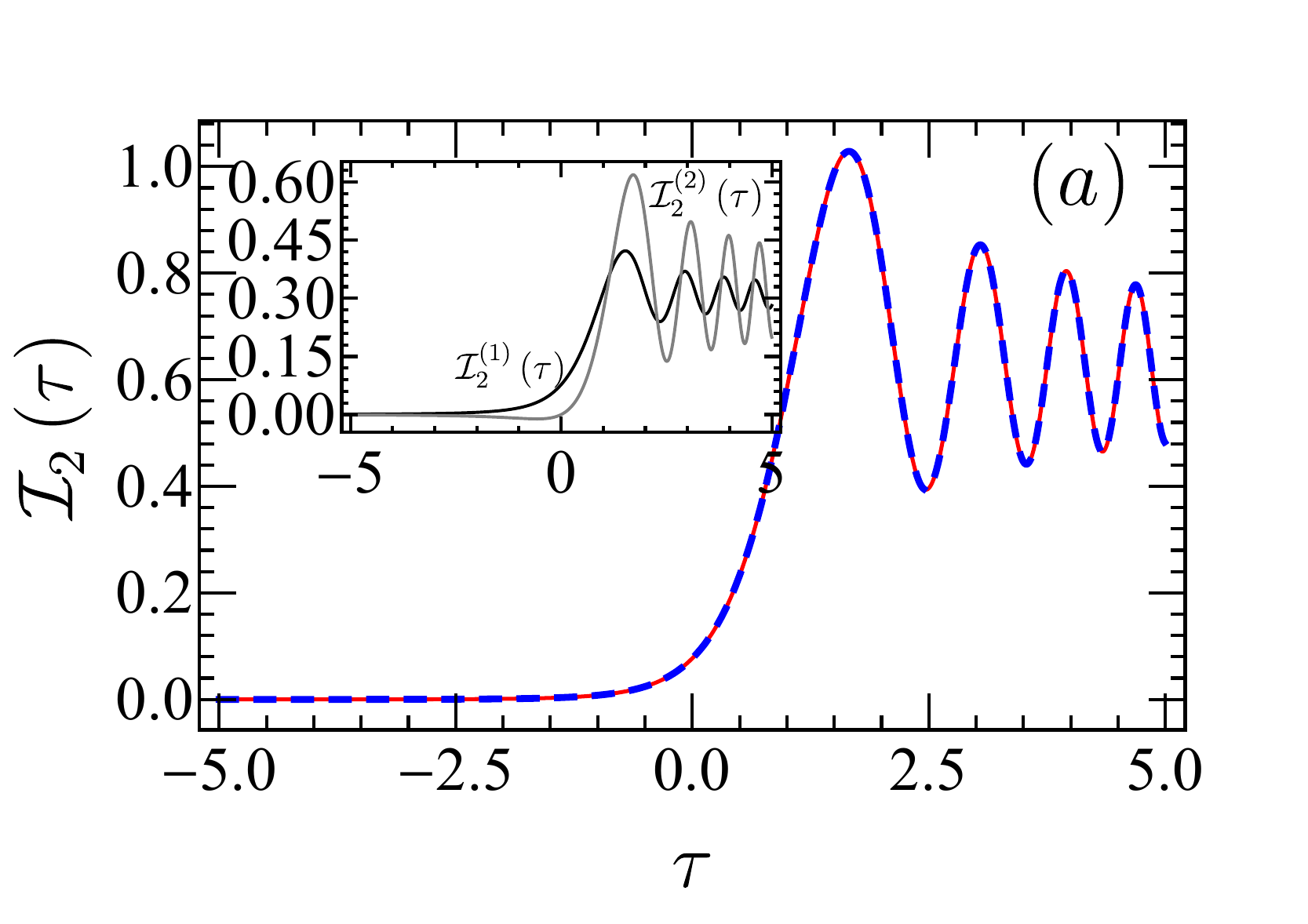}\hspace{-0.75cm}
			\includegraphics[width=8cm, height=5cm]{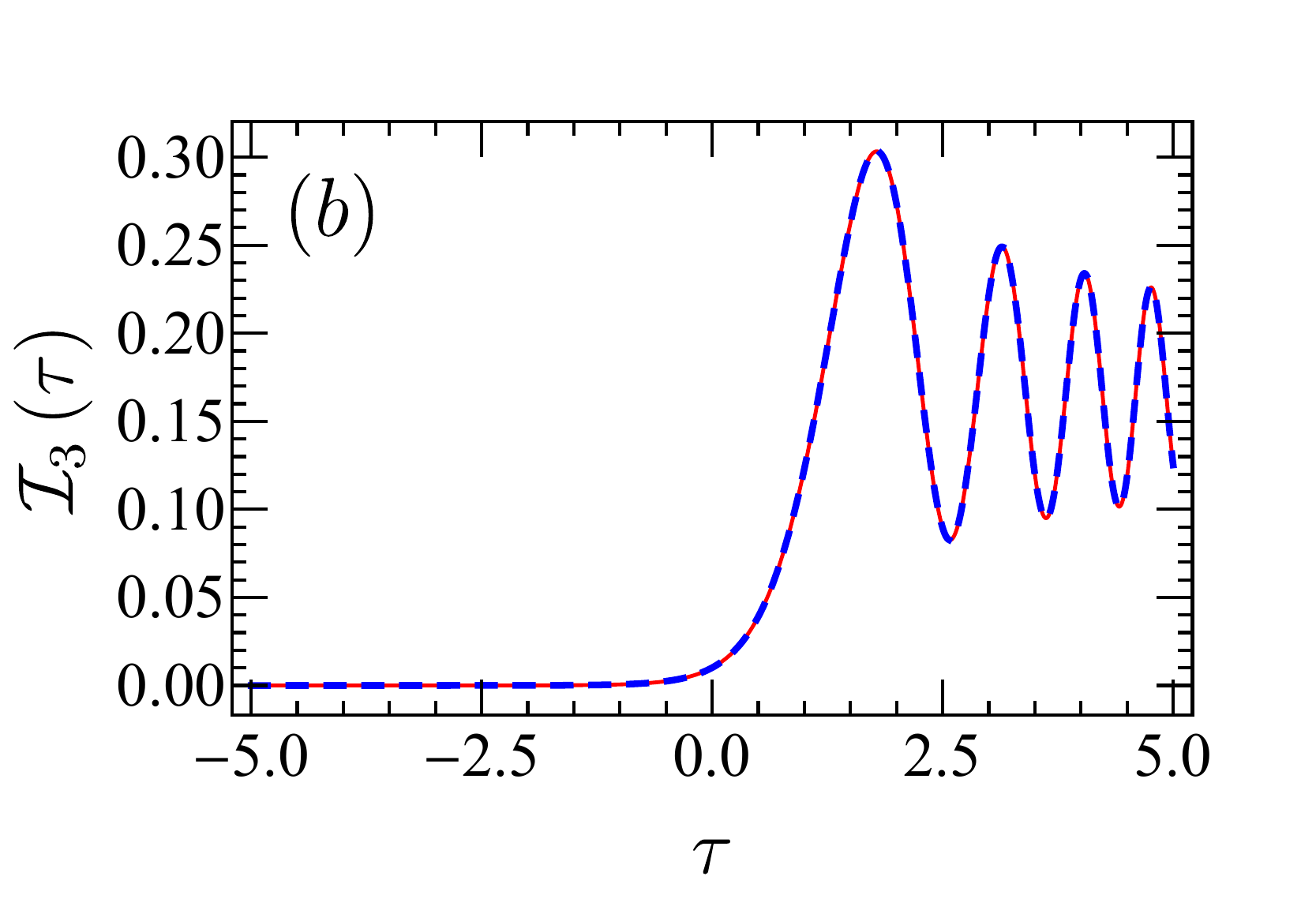}\\\vspace{-0.75cm}
			\includegraphics[width=8cm, height=5cm]{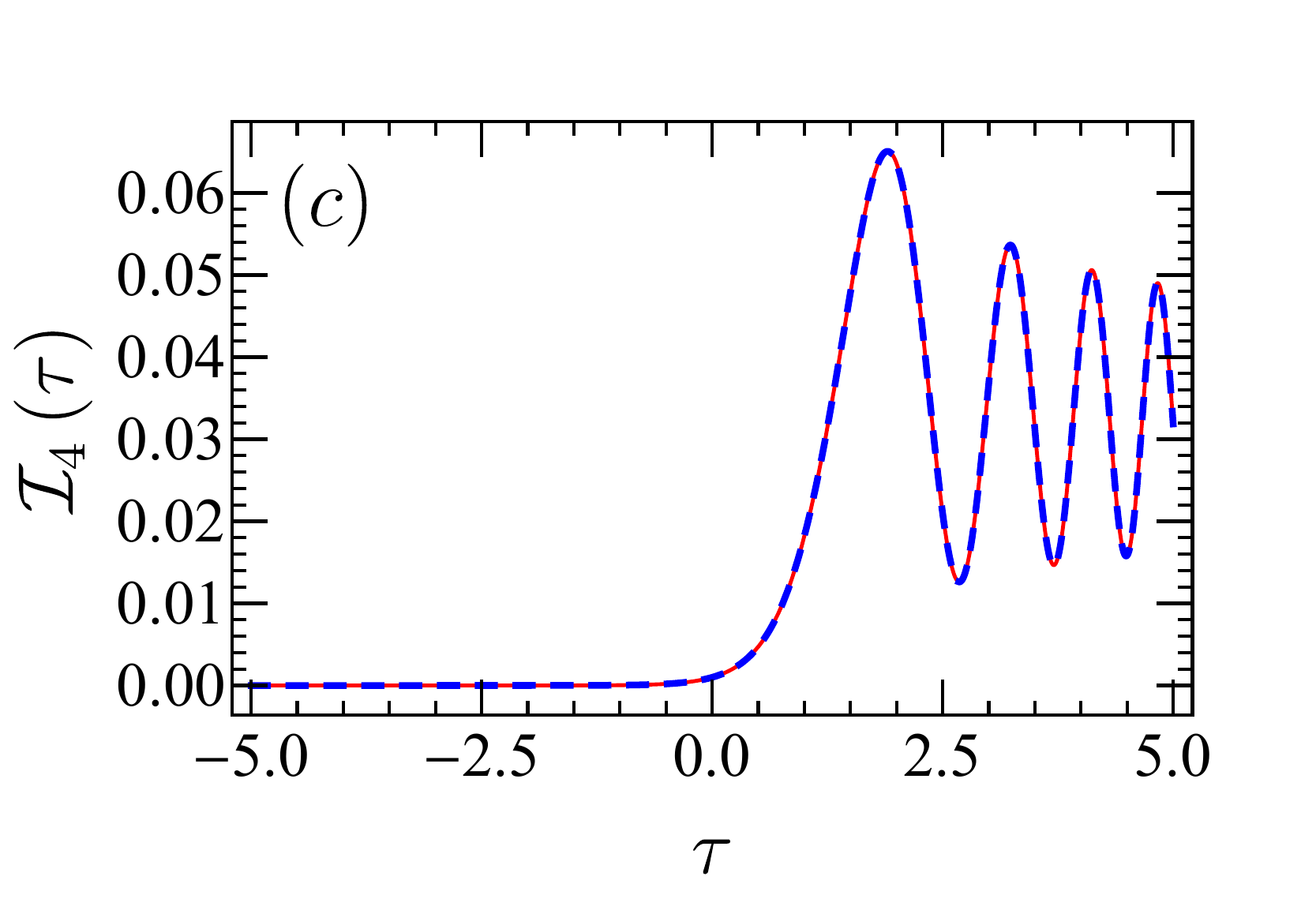}\hspace{-0.75cm}
			\includegraphics[width=8cm, height=5cm]{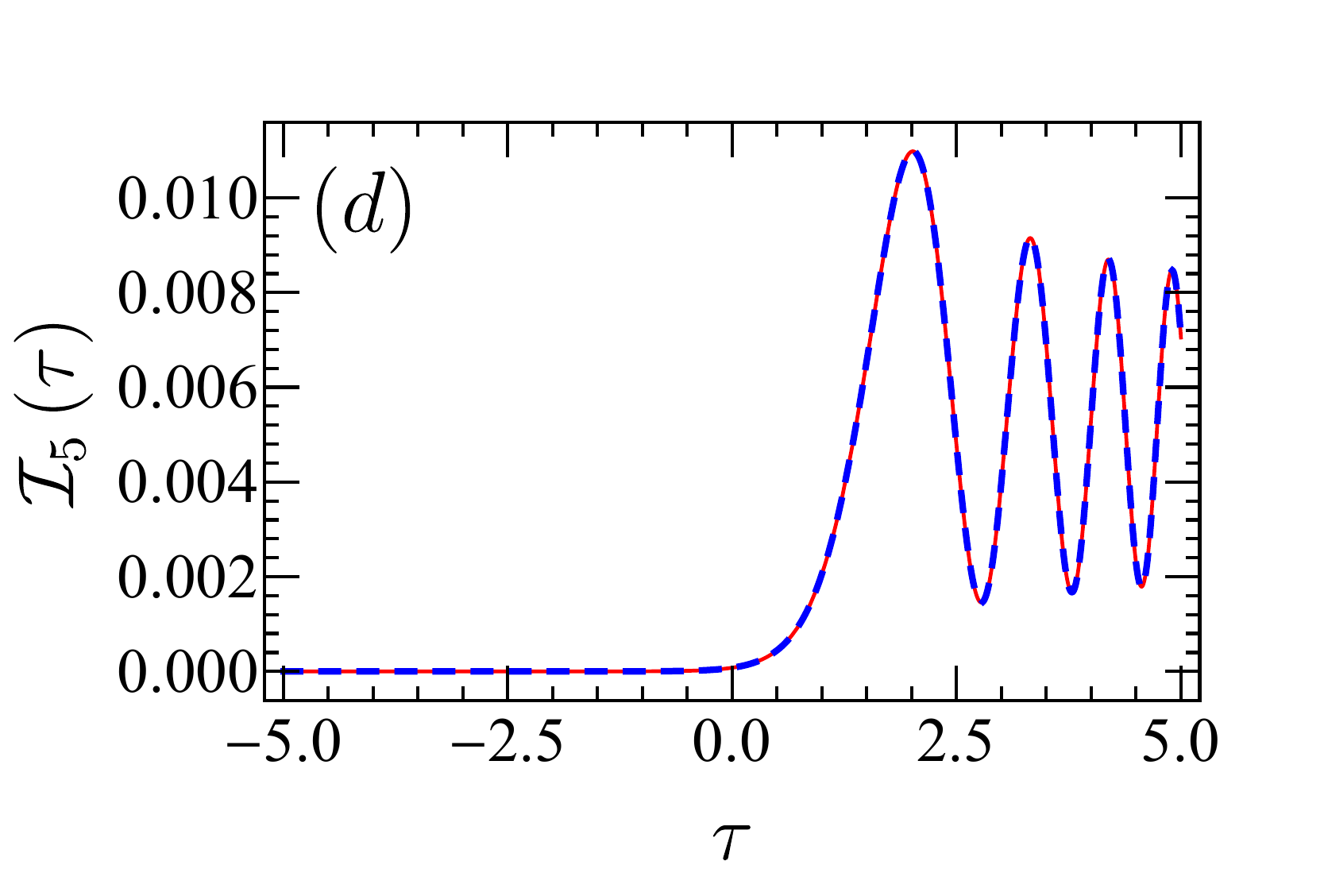}
			\vspace{-0.5cm}
			\caption{Benchmarking of the analytical formulas (blue dashed lines) against numerical simulations of the original integrals (solid red lines). For all integrals in the time window $]-\infty,0_{-}[$, almost nothing happens, the integrals are all nearly-zero. $\mathcal{I}_k(0)$ are given for arbitrary $k$ in Eq.\eqref{equ14b}. At the moment when the values of the integrals start becoming noticeable, one observes an abrupt jump of the integrals to their maximum values followed by St\"uckelberg oscillations with damping amplitudes. The duration of this jump is the transition time\cite{Vitanov1999,Yan}. We also observe that $\mathcal{I}_k(\tau)$ decreases with increasing $k$ confirming the existence of virtual transition paths.}
			\label{Figure2}
		\end{center}
	\end{figure}

	\subsection{Derivative of Parabolic Cylinder Weber function with respect to the index}
	As we have just mentioned, one of the main challenges encountered so far is inherent from the complexity related to the derivative of PCFs with respect to $\nu$. Fortunately, for the purpose of this work, we only need to know this for ${\rm Re}(\nu)<0$. We cope with this challenge by constructing an equivalent and simple expression. To this end, we make use of the following integral representation of PCFs\cite{Abramowitz, Fai2024},
	\begin{eqnarray}\label{equ16}
		D_{\nu}\left(z\right)=\frac{e^{-z^{2}/4}}{\Gamma(-\nu)}\int^{\infty}_0{dx}e^{-xz-x^{2}/2}x^{-\nu-1},\quad \mathrm{Re}\left(\nu\right)<0,
	\end{eqnarray} 
	where $\Gamma$ is the gamma function\cite{Abramowitz, Fai2024}.  Among all possible integral representations for  PCFs, Eq.\eqref{equ16} is convenient for our goal. However, its direct derivative with respect to $\nu$ involves a $\log$ function that renders our task more complicated. Instead, it is intuitive to perform a series expansion of the $z$ part of the kernel. Rigorously speaking, because, $z$ is a complex variable, a Taylor's series expansion about any point $z_0$ is not possible. This can only be done by using the Cauchy integral formula for derivatives,
	\begin{eqnarray}\label{equ17}
		D_{\nu}\left(z\right)=\frac{e^{-{z}^{2}/4}}{\Gamma(-\nu)}\sum^{\infty}_{n=0}{\frac{{(-z)}^{n}}{n!}}\int^{\infty}_0{dx}e^{-{x}^{2}/2}{x}^{-\nu-1+n},\quad \mathrm{Re}\left(\nu\right)<0.
	\end{eqnarray}  
	We transform the integral in the right-hand side of \eqref{equ17} with the change of variable $u=x^2$ and identify the integral representation of the $\Gamma$ function leading us to the adapted form
	\begin{eqnarray}\label{equ18}
		D_{\nu}\left(z\right)=\frac{e^{-{z}^{2}/4}}{\Gamma(-\nu)}\sum^{\infty}_{n=0}{\frac{{(-z)}^{n}}{n!}}{2}^{\frac{1}{2}(n-\nu-2)}\Gamma\left(\frac{n-\nu}{2}\right),\ \ \mathrm{Re}(\nu)<0,
	\end{eqnarray}  
	which can be regarded as a series representation of $D_{\nu}\left(z\right)$. Note that this is not the Taylor's series of $D_{\nu}\left(z\right)$. The series \eqref{equ18} absolutely converges for all complex $z$ in an open disc with an infinite radius. One can convince oneself by using the D'Alambert (ratio) criterion\cite{Abramowitz}. Indeed, let $a_n=\frac{(-z)^{n}}{n!}2^{n/2}\Gamma\left(\frac{n-\nu}{2}\right)$ be the term of the series and evaluate the ratio $|\frac{a_{n+1}}{a_n}|$. We prove that $|\frac{a_{n+1}}{a_n}|\sim \frac{n^{1/2}}{n+1}|z|$ whereby the D'Alambert ratio test $\lim_{n\to\infty}|\frac{a_{n+1}}{a_n}|\to 0$. This confirms that the series absolutely converges. For this proof, we have used the fact that $\Gamma(z+1/2)/\Gamma(z)\sim z^{1/2}+\mathcal{O}(z^{-1})$ when $z\to\infty$ and $(\frac{n-\nu}{2})^{1/2}\sim \left(\frac{n}{2}\right)^{1/2}+\mathcal{O}(n^{-1})$ when $n\to\infty$. The radius of convergence $R=\frac{1}{\lim\sup_{n\to\infty}|\frac{a_{n+1}}{a_n}|}=\infty$ as stated.  Therefore, \eqref{equ18} absolutely converges in the indicated domain.
	
	Keeping our main target in mind, we introduce the modified PCFs,
	\begin{eqnarray}\label{equ19}
		\mathcal{D}^{\left(k\right)}_{\nu}\left(z\right)=\frac{e^{-\frac{{z}^{2}}{4}}}{\Gamma\left(-\nu\right)}\sum^{\infty}_{n=0}{\frac{{\left(-z\right)}^{n}}{n!}}{2}^{\left(n-\nu-2\right)/2}\left(\frac{\partial^k \Gamma\left(\frac{n-\nu}{2}\right)}{\partial \nu^k}\right),\quad k\ge 0
	\end{eqnarray}
	which generalizes \eqref{equ18}. We have verified that it preserves certain properties of PCFs. Indeed, it is defined such that $\mathcal{D}^{\left(0\right)}_{p}\left(z\right)=D_{p}\left(z\right)$ and possesses the following derivative property $\frac{d}{dz}\mathcal{D}^{\left(k\right)}_{\nu}\left(z\right)=-\frac{z}{2}\mathcal{D}^{\left(k\right)}_{\nu}\left(z\right)+\nu \mathcal{D}^{\left(k\right)}_{\nu-1}\left(z\right)$ reminiscent to that of PCFs. Moreover, it is important to be equipped with the following relation,
	\begin{eqnarray}\label{equ20}
	\hspace{-1cm}	\frac{\partial^k }{\partial \nu^k}\Gamma\left(\frac{n-\nu}{2}\right)=\left(-\frac{1}{2}\right)^k\Gamma\left(\frac{n-\nu}{2}\right)\sum_{j=1}^{k}\mathcal{B}_{k,j}\left(\psi^{\left(0\right)},\psi^{\left(1\right)},\cdots,\psi^{\left(k-2\right)},\psi^{\left(k-1\right)}\right),\quad k>0
	\end{eqnarray}  
	where $\mathcal{B}_{k,j}\left(\cdots\right)$ is the partial Bell polynomials of order $k$ with $j$ partitions; $\psi^{\left(r\right)}\equiv \psi^{\left(r\right)}\left([n-\nu]/2\right)$ are Polygamma functions of order $r$ (see Refs.\onlinecite{Abramowitz, Fai2024}).
	It can therefore be inferred after some algebra that the desired derivative rule reads 
	\begin{eqnarray}\label{equ21}
		\frac{\partial \mathcal{D}^{\left(k\right)}_{\nu}\left(z\right)}{\partial \nu}=\phi_{\nu}\mathcal{D}^{\left(k\right)}_{\nu}\left(z\right)+\mathcal{D}^{\left(k+1\right)}_{\nu}\left(z\right),\quad \mathrm{Re}\left(\nu\right)<0,
	\end{eqnarray}  
	where
	\begin{eqnarray}\label{equ22}
		\phi_{\nu}=\psi^{\left(0\right)}\left(-\nu\right)-\frac{\ln 2}{2}.
	\end{eqnarray} 
	This is much simpler and easier to handle compared to the expression in Ref.\onlinecite{Mathematica}. It should also be noted that Ref.\onlinecite{Mathematica} is the only source where we have found the derivative of PCFs with respect to $\nu$. Eq.\eqref{equ21} has the form the derivative of PCF is expected to have. By virtue of \eqref{equ21}, higher order derivatives can also be expressed in closed-form through a simple expression as,
	\begin{eqnarray}\label{equ23}
		\frac{\partial^n \mathcal{D}^{\left(k\right)}_{\nu}\left(z\right)}{\partial \nu^n}=\sum_{m=0}^{n}\mathcal{P}_{n,m}\left(\nu\right)\mathcal{D}^{\left(k+m\right)}_{\nu}\left(z\right),\quad \mathrm{Re}\left(\nu\right)<0,
	\end{eqnarray} 
	where
	\begin{eqnarray}\label{equ24}
		\mathcal{P}_{n,m}\left(\nu\right) = \frac{n!}{m!\left(n-m\right)!}\hat{T}_{\nu}^{n-m-1}\phi_{\nu}, \quad n\neq m,
	\end{eqnarray} 
	and
	\begin{eqnarray}\label{equ25}
		\hat{T}_{\nu} = \frac{\partial}{\partial\nu} + \phi_{\nu},
	\end{eqnarray}
	with $\mathcal{P}_{n,m}\left(\nu\right)=0$ for $ n<m$ or $m<0$ and $\mathcal{P}_{n,n}\left(\nu\right)=1$. Note that Eq.\eqref{equ21} is exactly of the form \eqref{equ23} for $n=1$ with  $\mathcal{P}_{1,0}\left(\nu\right)=\phi_\nu$ and  $\mathcal{P}_{1,1}\left(\nu\right)=1$. Having the above tools at hand, we are now in position to cope with our main target. We invoke once again the Leibniz formula with the help of which we attain,
	\begin{eqnarray}\label{equ26}
	\nonumber	\frac{\partial^n }{\partial \nu^n}\Big|\mathcal{D}^{\left(r\right)}_{-i\nu-1}\left(-i\mu_0\tau\right)\Big|^2=\sum_{j=0}^{n}\sum_{k=0}^{n-j}\sum_{\ell=0}^{j}\binom{n}{k}\mathcal{P}_{n-j,k}\left(\nu\right)\mathcal{P}^{*}_{j,\ell}\left(\nu\right)\mathcal{D}^{\left(r+k\right)}_{-i\nu-1}\left(-i\mu_{0}\tau\right)\mathcal{D}^{\left(r+\ell\right)}_{i\nu-1}\left(i\bar{\mu}_0\tau\right),\\
	\end{eqnarray} 
	where
	\begin{eqnarray}\label{equ27}
		\mathcal{P}_{n,m}\left(\nu\right) = \frac{n!(-i)^{m}}{m!\left(n-m\right)!}\hat{T}_{\nu}^{n-m-1}\phi^{\left(1\right)}\left(\nu\right).
	\end{eqnarray} 
	Upon acting the operator $\hat{T}_{\nu} = \left(\partial/\partial\nu + \phi^{\left(1\right)}_{\nu}\right)$ $n$-times onto the function $\phi^{\left(1\right)}\left(\nu\right)\equiv -i\phi_{-i\nu-1}$ yields,
	\begin{eqnarray}\label{equ27a}
		\hat{T}_{\nu}^{n}\phi^{\left(1\right)}\left(\nu\right)=\sum_{j=1}^{n+1}\mathcal{B}_{n+1,j}\left(\phi^{\left(1\right)},\phi^{\left(2\right)},\cdots,\phi^{\left(n\right)},\phi^{\left(n+1\right)}\right),\quad n\ge 0
	\end{eqnarray} 
	where $\phi^{\left(r\right)}\left(\nu\right)\equiv-i^{r}\psi^{\left(r-1\right)}\left(i\nu+1\right)$ for $r>1$. Here as before, $\mathcal{P}_{n,m}\left(\nu\right)=0$ for $ n<m$ or $m<0$ and $\mathcal{P}_{n,n}\left(\nu\right)=(-i)^n$.
	Letting $\nu=0$ in \eqref{equ26} leads us to our main result \eqref{equ7} achieving our goal. 
	At long positive time where $P_{\rm LZ}(\infty,\nu)=1-e^{-2\pi\nu}$ one has 
	\begin{eqnarray}\label{equ14}
		\frac{\partial^{k}P_{\rm LZ}(\infty,\nu)}{\partial \nu^{k}}=-(-2\pi)^{k}e^{-2\pi\nu}, 
	\end{eqnarray} 
	leading to the Kholodenko-Silagadze result \eqref{equ2}. At level-crossing, the transition and occupation probabilities are respectively given by  $P_{\rm LZ}(0,\nu)=\frac{1}{2}\left(1-e^{-\pi\nu}\right)$ and $1-P_{\rm LZ}(0,\nu)=\frac{1}{2}\left(1+e^{-\pi\nu}\right)$. In the strong adiabtic limit $\nu\to\infty$ both diabatic levels have occupancies equal to $1/2$. From here,
	\begin{eqnarray}\label{equ14a}
		\frac{\partial^{k}P_{\rm LZ}(0,\nu)}{\partial \nu^{k}}=-\frac{1}{2}\left(-\pi\right)^{k}e^{-\pi\nu}, 
	\end{eqnarray} 
	and
	\begin{eqnarray}\label{equ14b}
		\mathcal{I}_{k}\left(0\right)=\frac{\pi^{k}}{2^{3k}k!}.
	\end{eqnarray}
	We also verify that $\mathcal{I}_{0}\left(0\right)=1$. 

	\subsection{Numerical Analysis}\label{Sec2b}
	Due to the lack of Mathematical tools for a direct proof of Eq.\eqref{equ5}, we can only support our assertions with data from numerical simulations. We start painting out our  approach with  $\mathcal{I}_{1}(\tau)$. It should be indicated that as a nested double integral with highly oscillating integrand, $\mathcal{I}_{1}(\tau)$ is numerically expensive. It is highly unstable and for this reason, built-in functions for numerical integrations in symbolic calculators can be misleading. We avoid this issue by following Levin's arguments\cite{Chen2024}. Thus, by successively working out the time-derivative of $\mathcal{I}_{1}(\tau)$ while eliminating any occurrence of the sine and cosine functions transforms it into the following third order linear ODE,
	\begin{eqnarray}\label{equ7b}
		\frac{d^3}{d\tau^3}\mathcal{I}_{1}(\tau)-\frac{1}{\tau}\frac{d^2}{d\tau^2}\mathcal{I}_{1}(\tau)+4\tau^2\frac{d}{d\tau}\mathcal{I}_{1}(\tau)=-\frac{1}{\tau},
	\end{eqnarray}
	which is solved with initial conditions 
	\begin{eqnarray}\label{equ7c}
		\frac{d^2}{d\tau^2}\mathcal{I}_{1}(\tau)|_{\tau\to-\infty}=1,\quad \frac{d}{d\tau}\mathcal{I}_{1}(\tau)|_{\tau\to-\infty}=\mathcal{I}_{1}(-\infty)=0.
	\end{eqnarray}
	This is stable, robust and converges faster with higher precision compared to its double integral counterpart. Eq.\eqref{equ7b} exactly reproduces Eq.(24) in Ref.\onlinecite{Kenmoe2013} for $\lambda=0$ and our solution coincides with Eq.(26) under the same condition. These remarks validate our analytical approach. Because the homogeneous part of \eqref{equ7b} can be reduced to a second order ODE and exactly solved, a particular solution can be obtained through the Green function method. However, this falls out of the scope of the present interests. Based on the numerical approach, we could visualize $\mathcal{I}_{1}(\tau)$. Its gross temporal profile, a Fourier transform and a parametric plot are displayed in Figure \ref{Figure0}.
	
	In general, the above technique for arbitrary $k$ leads to non-homogeneous $(2k+1)$th order linear ODEs. For instance, for $k=2$, we have a fifth order ODE. One immediately sees that for $k>2$, the task will become more difficult and even impossible to handle.  Instead, we exploit the fact that $\mathcal{I}_{k}(\tau)$ and $\mathcal{I}_{k-1}(\tau)$ are connected through the double integral relation,
	\begin{eqnarray}\label{equ7d}
		\mathcal{I}_{k}(\tau)=\int^\tau_{-\infty}d\tau_{1}\int^{\tau_{1}}_{-\infty}d\tau_{2}\cos[\tau^{2}_{1}-\tau^{2}_{2}]\mathcal{I}_{k-1}\left(\tau_2\right), \quad k>0
	\end{eqnarray}
	with $\mathcal{I}_{0}(\tau)=1$. This integral is similar to that for $\mathcal{I}_{1}(\tau)$ as described above and can therefore be transformed into a non-homogeneous third order linear ODE. This is done in a similar fashion by computing successive time derivatives of \eqref{equ7d} and eliminating any occurrence of sine and cosine functions. The procedure yields the ODE,
	\begin{eqnarray}\label{equ7e}
		\frac{d^3}{d\tau^3}\mathcal{I}_{k}(\tau)-\frac{1}{\tau}\frac{d^2}{d\tau^2}\mathcal{I}_{k}(\tau)+4\tau^2\frac{d}{d\tau}\mathcal{I}_{k}(\tau)=\frac{d}{d\tau}\mathcal{I}_{k-1}(\tau)-\frac{1}{\tau}\mathcal{I}_{k-1}(\tau),
	\end{eqnarray}
	with initial conditions
	\begin{eqnarray}\label{equ7f}
		\frac{d^2}{d\tau^2}\mathcal{I}_{k}(\tau)|_{\tau\to-\infty}=\mathcal{I}_{k-1}(-\infty),\quad \frac{d}{d\tau}\mathcal{I}_{k}(\tau)|_{\tau\to-\infty}=\mathcal{I}_{k}(-\infty)=0.
	\end{eqnarray}
	The values of $\mathcal{I}_{k-1}(\tau)$ serve as source terms and initial conditions for $\mathcal{I}_{k}(\tau)$. 
	Numerical simulations become more robust and stable by transforming again the third order ODE into a system of three first order linear ODE. This is done by defining $y_1\left(\tau\right)=\frac{d}{d\tau}\mathcal{I}_{k}(\tau)$, $y_2\left(\tau\right)=\frac{d^2}{d\tau^2}\mathcal{I}_{k}(\tau)$ and $y_3\left(\tau\right)=\frac{d^3}{d\tau^3}\mathcal{I}_{k}(\tau)$ leading to
	\begin{eqnarray}\label{equ7g}
		\nonumber\frac{d}{d\tau}y_1\left(\tau\right)=y_2\left(\tau\right),\quad \frac{d}{d\tau}y_2\left(\tau\right)=y_3\left(\tau\right),\\ \frac{d}{d\tau}y_3\left(\tau\right)=\frac{1}{\tau}y_3\left(\tau\right)-4\tau^2y_2\left(\tau\right)+\left(\frac{d}{d\tau}\mathcal{I}_{k-1}(\tau)-\frac{1}{\tau}\mathcal{I}_{k-1}(\tau)\right),
	\end{eqnarray}
	with initial conditions 
	\begin{eqnarray}\label{equ7h}
		y_1\left(-\infty\right)=y_2\left(-\infty\right)=0,\quad \text{and}\quad y_3\left(-\infty\right)=\mathcal{I}_{k-1}(-\infty).
	\end{eqnarray}
	This technique has allowed us to verify $\mathcal{I}_{k}(\tau)$ numerically for $k=\{2,3,4,5\}$ (See Figure.\ref{Figure2}). This figure shows that $\mathcal{I}_k(\tau)$ rapidly decreases when $k$ increases. From $\mathcal{I}_{2}(\tau)$ to $\mathcal{I}_{5}(\tau)$ the value of the integrals drop by $99\%$. This results from destructive interference between virtual paths. We also observe that the transition time is different for each integral. 
	
	\section{Illustrative Examples and more results}\label{Sec4}
	\subsection{Illustrative Examples}\label{Sec4.1}
	It is important at this stage to illustrate the above closed-form calculations with explicit and concrete examples. Thus, from \eqref{equ15} we manually infer a few order derivatives,
	\begin{eqnarray}\label{equ28}
		\frac{\partial P_{\rm LZ}\left(\tau,\nu\right)}{\partial\nu}\Big|_{\nu=0}=\Big[{\left|D_{-i\nu-1}\left(-i\mu_0\tau\right)\right|}^{2}\Big]_{\nu=0},
	\end{eqnarray}  
	\begin{eqnarray}\label{equ29}
		\frac{{\partial}^{2}P_{\rm LZ}(\tau,\nu)}{\partial{\nu}^{2}}\Big|_{\nu=0}=\Big[-\pi{\left|D_{-i\nu-1}\left(-i\mu_0\tau\right)\right|}^{2}\mathrm{+2}\frac{\partial{\left|D_{-i\nu-1}\left(-i\mu_0\tau\right)\right|}^{2}}{\partial\nu}\Big]_{\nu=0},
	\end{eqnarray}  
	\begin{eqnarray}\label{equ30}
		\frac{{\partial}^{3}P_{\rm LZ}\left(\tau,\nu\right)}{\partial{\nu}^{3}}\Big|_{\nu=0}=\Big[\frac{3\pi^{2}}{4}{\left|D_{-1}\left(-i\mu_0\tau\right)\right|}^{2}-3\pi\frac{\partial{\left|D_{-i\nu-1}\left(-i\mu_0\tau\right)\right|}^{2}}{\partial\nu}\mathrm{+3}\frac{{\partial}^{2}{\left|D_{-i\nu-1}\left(-i\mu_0\tau\right)\right|}^{2}}{\partial{\nu}^{2}}\Big]_{\nu=0},
	\end{eqnarray}  
	\begin{eqnarray}\label{equ31}
		\nonumber \frac{{\partial}^{4}P_{\rm LZ}(\tau,\nu)}{\partial{\nu}^{4}}\Big|_{\nu=0}&=&\Big[-\frac{\pi^{3}}{2}{\left|D_{-i\nu-1}\left(-i\mu_0\tau\right)\right|}^{2}+3\pi^{2}\frac{\partial{\left|D_{-i\nu-1}\left(-i\mu_0\tau\right)\right|}^{2}}{\partial\nu}\\&-&6\pi\frac{{\partial}^{2}{\left|D_{-i\nu-1}\left(-i\mu_0\tau\right)\right|}^{2}}{\partial{\nu}^{2}}+4\frac{{\partial}^{3}{\left|D_{-i\nu-1}\left(-i\mu_0\tau\right)\right|}^{2}}{\partial{\nu}^{3}}\Big]_{\nu=0}
	\end{eqnarray}
	and
	\begin{eqnarray}\label{equ32}
		\nonumber \frac{{\partial}^{5}P_{\rm LZ}(\tau,\nu)}{\partial{\nu}^{5}}\Big|_{\nu=0}&=&\Big[\frac{5\pi^{4}}{16}\left|D_{-i\nu-1}\left(-i\mu_0\tau\right)\right|^{2}-\frac{5\pi^{3}}{2}\frac{{\partial}{\left|D_{-i\nu-1}\left(-i\mu_0\tau\right)\right|}^{2}}{\partial{\nu}}\\\nonumber&+&\frac{15\pi^{2}}{2}\frac{{\partial}^{2}{\left|D_{-i\nu-1}\left(-i\mu_0\tau\right)\right|}^{2}}{\partial{\nu}^{2}}-10\pi\frac{{\partial}^{3}{\left|D_{-i\nu-1}\left(-i\mu_0\tau\right)\right|}^{2}}{\partial{\nu}^{3}}\\&+&5\frac{{\partial}^{5}{\left|D_{-i\nu-1}\left(-i\mu_0\tau\right)\right|}^{2}}{\partial{\nu}^{5}}\Big]_{\nu=0}.
	\end{eqnarray}  
	The above relations partially determine the integrals $\mathcal{I}_{1}(\tau)$, $\mathcal{I}_{2}(\tau)$, $\mathcal{I}_{3}(\tau)$, $\mathcal{I}_{4}(\tau)$ and $\mathcal{I}_{5}(\tau)$ at finite times. We developed a separate Mathematica package called {\bf DerivativeOfModulusOfPCF.wl} available in our  \href{https://github.com/Kenmax15/Closed-Form-Approach-to-Oscillatory-Integrals/tree/main}{GitHub repository} for the symbolic computations of the derivatives without further calculations. In order to validate the outputs of this package the desired derivatives are manually calculated and used for benchmarking. 
	
	\subsubsection{First order derivative}
	The derivative $\partial\Big|\mathcal{D}^{\left(k\right)}_{-i\nu-1}\left(-iz\right)\Big|^2/\partial \nu$ can directly be calculated from \eqref{equ26}. We recall from \eqref{equ23} that 
	\begin{eqnarray}\label{equ32a}
		\frac{\partial \mathcal{D}^{\left(k\right)}_{-i\nu-1}\left(-i\mu_0\tau\right)}{\partial \nu}=\sum_{m=0}^{1}\mathcal{P}_{1,m}\left(\nu\right)\mathcal{D}^{\left(k+m\right)}_{-i\nu-1}\left(-i\mu_0\tau\right),
	\end{eqnarray} 
	where $\mathcal{P}_{1,0}(\nu)=-i\phi_{-i\nu-1}$ and $\mathcal{P}_{1,1}(\nu)=-i$. Taking the complex conjugate of \eqref{equ32a} and calculating the product $\mathcal{D}^{\left(k\right)}_{-i\nu-1}\left(-i\mu_0\tau\right)\mathcal{D}^{\left(k\right)}_{i\nu-1}\left(i\bar{\mu}_0\tau\right)$ leads us after a few manipulations to 
	\begin{eqnarray}\label{equ33}
	\nonumber	\frac{\partial\left|\mathcal{D}^{\left(k\right)}_{-i\nu-1}\left(-i\mu_{0}\tau\right)\right|^{2}}{\partial\nu}=2\mathrm{Im}\left(n_{-i\nu-1}\right){\left|\mathcal{D}^{\left(k\right)}_{-i\nu-1}\left(-i\mu_{0}\tau\right)\right|}^{2}-2\mathrm{Im}\left[\mathcal{D}^{\left(k\right)}_{-i\nu-1}\left(-i\bar{\mu}_{0}\tau\right)\mathcal{D}_{-i\nu-1}^{\left(k+1\right)*}\left(-i\bar{\mu}_{0}\tau\right)\right],\\
	\end{eqnarray}
	whereby setting $\nu=0$ reveals the obvious fact that $\mathrm{Im}\left(\phi_{-1}\right)=0$ yielding
	\begin{eqnarray}\label{equ32b}
		\frac{\partial\left|\mathcal{D}^{\left(k\right)}_{-i\nu-1}\left(-i\mu_{0}\tau\right)\right|^{2}}{\partial\nu}\Big|_{\nu=0}=-2\mathrm{Im}\left[\mathcal{D}^{\left(k\right)}_{-1}\left(-i\mu_{0}\tau\right)\mathcal{D}_{-1}^{\left(k+1\right)}\left(i\bar{\mu}_{0}\tau\right)\right].
	\end{eqnarray}
	Plugging \eqref{equ32b} with $k=0$ into \eqref{equ29} while considering \eqref{equ13} yields $\mathcal{I}_{2}(\tau)$ as displayed in \eqref{equ4}.
	
	\subsubsection{Second order derivative}
\begin{figure}[]
		\vspace{-0.5cm}
		\centering
		\begin{center} 
			\includegraphics[width=8cm, height=5cm]{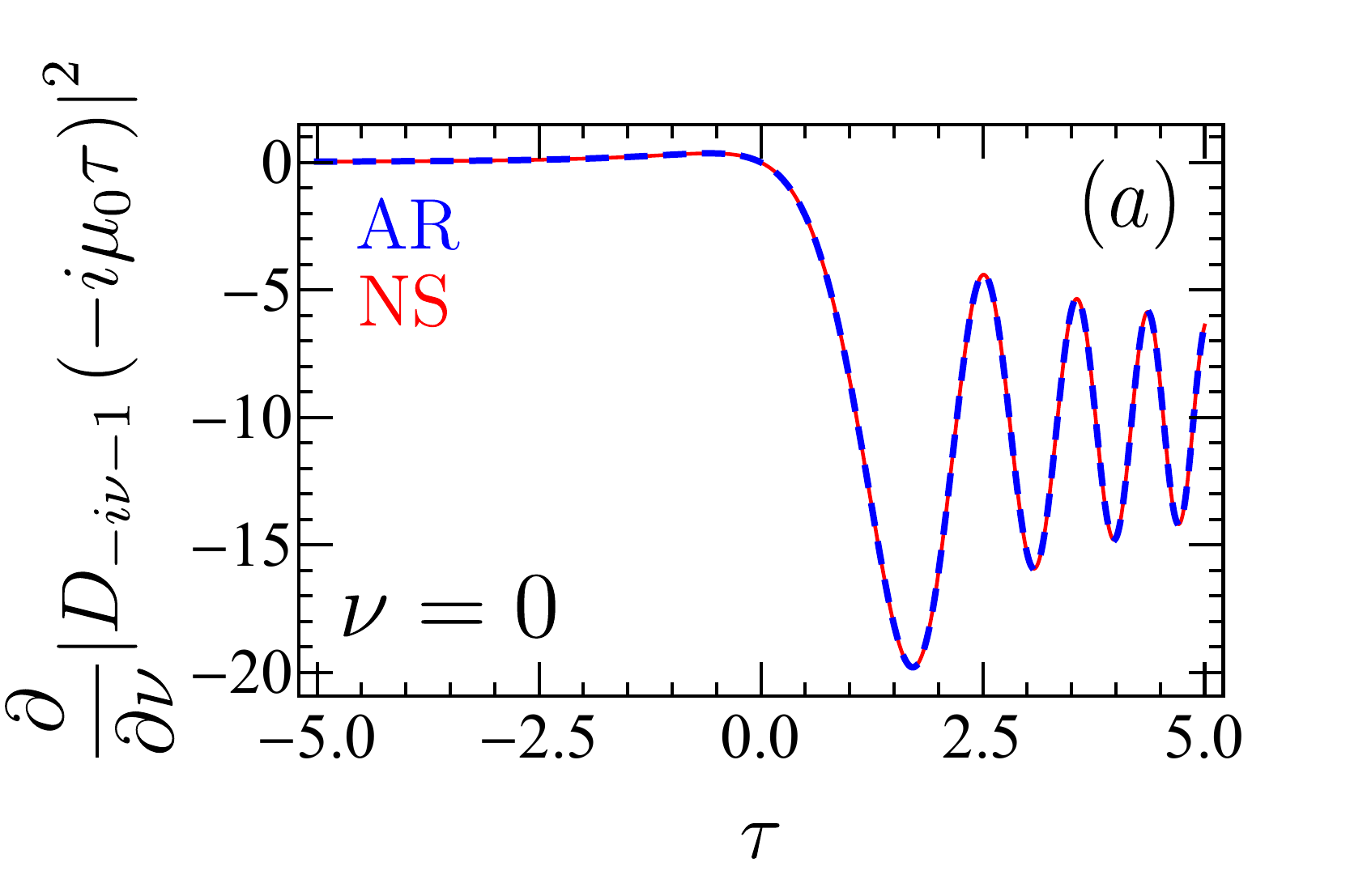}\hspace{-0.5cm}
			\includegraphics[width=8cm, height=5cm]{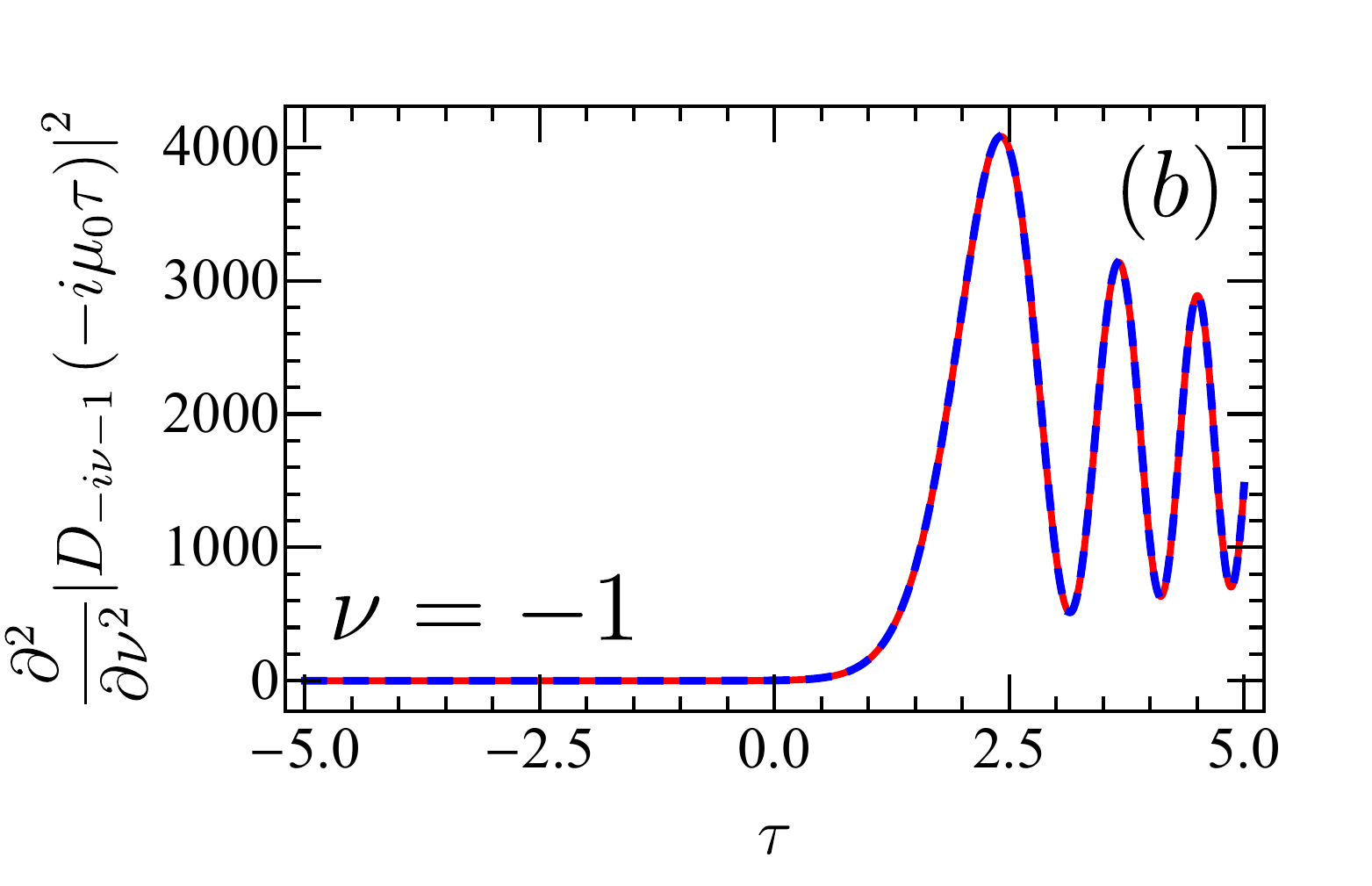}\\\vspace{-0.75cm}
			\includegraphics[width=8cm, height=5cm]{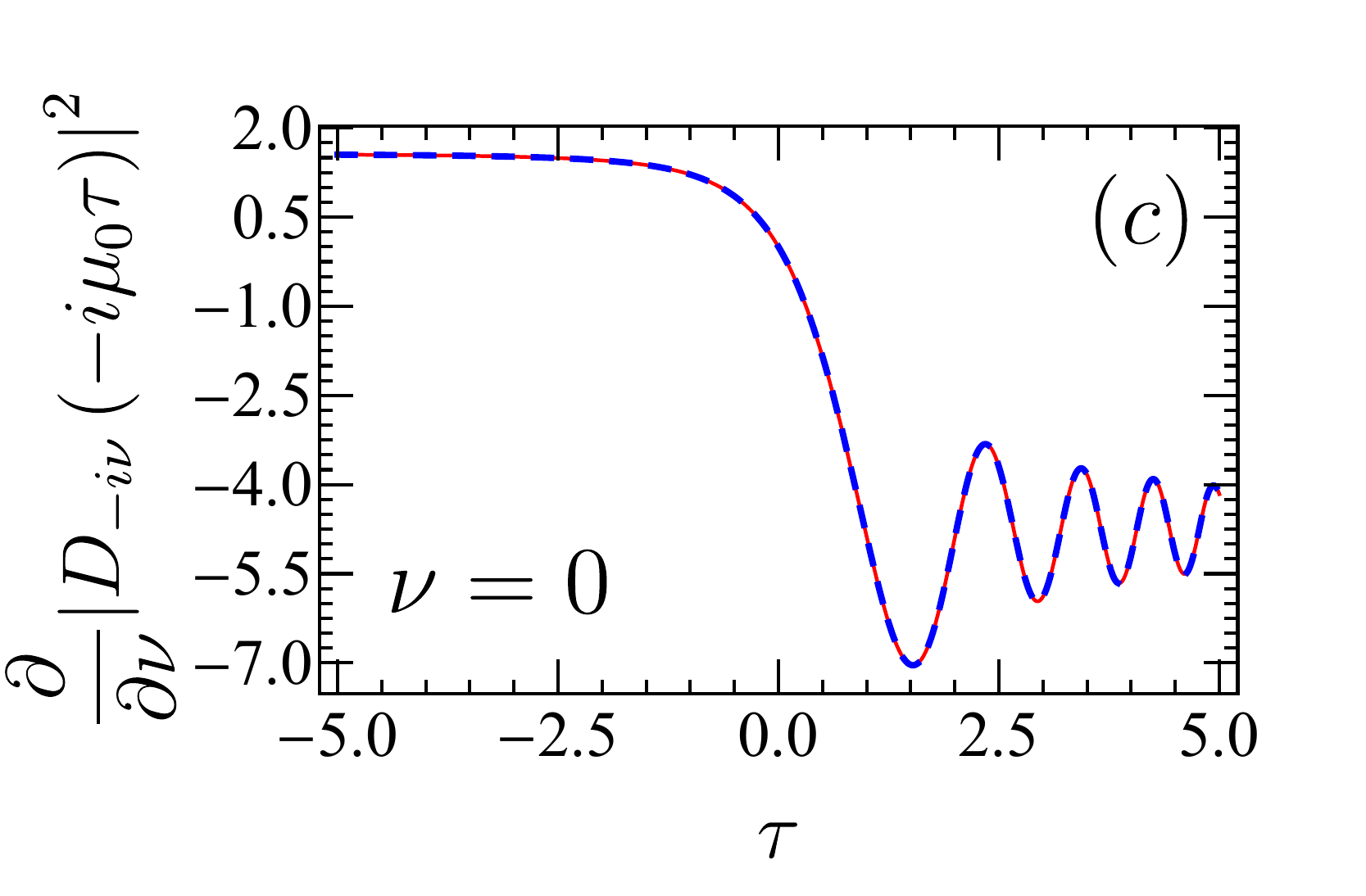}\hspace{-0.5cm}
			\includegraphics[width=8cm, height=5cm]{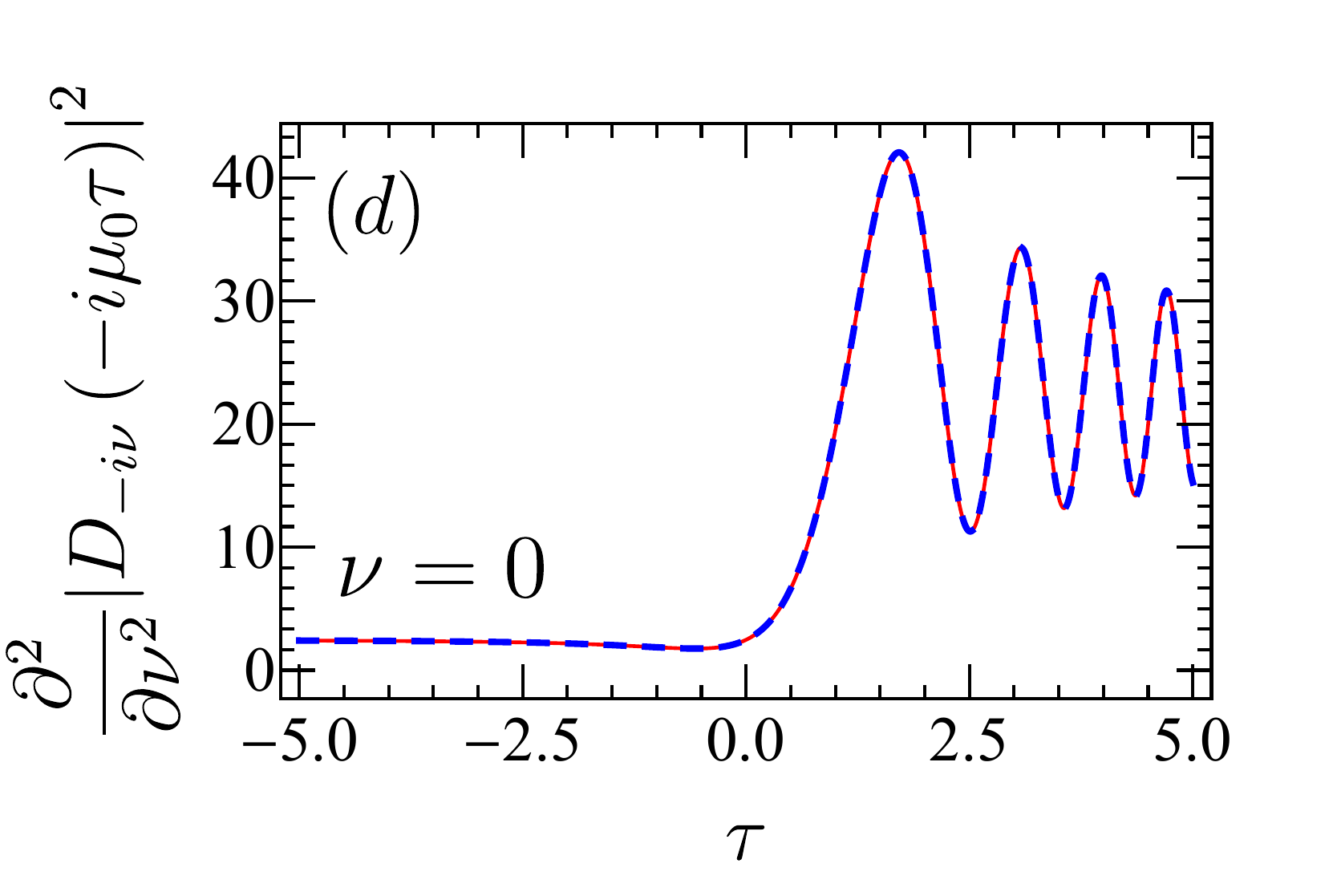}
			\vspace{-0.5cm}
			\caption{Legend: \color{blue}AR\color{black} (Analytical result); \color{red}NS\color{black} (Numerical result). Comparison between the data obtained by numerical simulations of the original derivatives (Red solid lines) and our analytical formula (\ref{equ32b}) and \eqref{equ32c} (Dashed blue lines). The perfect agreement between both results is an indication that our analytical calculations are correct.}
			\label{Figure1}
		\end{center}
	\end{figure}

	The second order derivative is calculated in a similar fashion as we did for the first order. After lengthy but straightforward calculations, we obtain 
	\begin{eqnarray}\label{equ32c}
	\nonumber	\frac{\partial^2\left|\mathcal{D}^{\left(k\right)}_{-i\nu-1}\left(-i\mu_{0}\tau\right)\right|^{2}}{\partial\nu^2}\Big|_{\nu=0}=\frac{\pi^2}{3}\left|\mathcal{D}^{\left(k\right)}_{-1}\left(-i\mu_{0}\tau\right)\right|^{2}-2\mathrm{Re}\left[\mathcal{D}^{\left(k\right)}_{-1}\left(-i\mu_{0}\tau\right)\mathcal{D}_{-1}^{\left(k+2\right)}\left(i\bar{\mu}_{0}\tau\right)\right]\\+2\left|\mathcal{D}^{\left(k+1\right)}_{-1}\left(-i\mu_{0}\tau\right)\right|^{2}.
	\end{eqnarray}
	The package in our \href{https://github.com/Kenmax15/Closed-Form-Approach-to-Oscillatory-Integrals/tree/main}{GitHub repository} gives exactly the same result. Numerical calculations allowed us to verify our analytical derivations. Figure \ref{Figure1} compares the data from \eqref{equ32c} and the results from numerical simulations. We observe a perfect agreement between both data. Eq.\eqref{equ32c} with the first order derivative \eqref{equ32b} completely determines $\mathcal{I}_{3}(\tau)$.
	\subsubsection{Third order derivative}
	After a certain effort,
	\begin{eqnarray}\label{equ32d}
		\nonumber\frac{\partial^3\left|\mathcal{D}^{\left(k\right)}_{-i\nu-1}\left(-i\mu_{0}\tau\right)\right|^{2}}{\partial\nu^3}\Big|_{\nu=0}=-2\pi^2\mathrm{Im}\left[\mathcal{D}^{\left(k\right)}_{-1}\mathcal{D}_{-1}^{\left(k+1\right)*}\right]-6\mathrm{Im}\left[\mathcal{D}^{\left(k+1\right)}_{-1}\mathcal{D}_{-1}^{\left(k+2\right)*}\right]+2\mathrm{Im}\left[\mathcal{D}^{\left(k\right)}_{-1}\mathcal{D}_{-1}^{\left(k+3\right)*}\right],\\
	\end{eqnarray}
	where we have used the shorthand notations $\mathcal{D}^{\left(k\right)}_{-1}\equiv\mathcal{D}^{\left(k\right)}_{-1}\left(-i\mu_{0}\tau\right)$ and $\mathcal{D}^{\left(k\right)*}_{-1}\equiv\mathcal{D}^{\left(k\right)}_{-1}\left(i\bar{\mu}_{0}\tau\right)$. Our package rigorously outputs the same result fully determining $\mathcal{I}_{4}(\tau)$.
	
	\subsubsection{Fourth order derivative}
	Likewise,
	\begin{eqnarray}\label{equ32e}
		\nonumber\frac{\partial^4\left|\mathcal{D}^{\left(k\right)}_{-i\nu-1}\left(-i\mu_{0}\tau\right)\right|^{2}}{\partial\nu^4}\Big|_{\nu=0}&=&\frac{\pi^4}{5}\Big|\mathcal{D}^{\left(k\right)}_{-1}\Big|^2+4\pi^2\Big|\mathcal{D}^{\left(k+1\right)}_{-1}\Big|^2-4\pi^2\mathrm{Re}\left[\mathcal{D}^{\left(k\right)}_{-1}\mathcal{D}_{-1}^{\left(k+2\right)*}\right]\\&-&8\mathrm{Re}\left[\mathcal{D}^{\left(k+1\right)}_{-1}\mathcal{D}_{-1}^{\left(k+3\right)*}\right]+2\mathrm{Re}\left[\mathcal{D}^{\left(k\right)}_{-1}\mathcal{D}_{-1}^{\left(k+4\right)*}\right]+6\Big|\mathcal{D}^{\left(k+2\right)}_{-1}\Big|^2,
	\end{eqnarray}
	determining $\mathcal{I}_{5}(\tau)$. Same as the result from package. These derivations validate all our packages as all analytical formula manually obtained reproduce the result from symbolic calculations. Additional numerical simulations, done but not shown, are used for benchmarking validating our analytical approach.
	
	\subsubsection{Results}
	The expressions for $\mathcal{I}_{4}(\tau)$ and $\mathcal{I}_{5}(\tau)$ are presented. Their simplified forms are achieved after considering Eqs.\eqref{equ32a}-\eqref{equ32e}
	\begin{eqnarray}\label{equ32f}
		\nonumber\mathcal{I}_{4}(\tau)&=&\frac{1}{2^{11}. 4!}\Big[-\frac{5\pi^3}{2}\left|D_{-1}\left(-i\mu_{0}\tau\right)\right|^{2}-12\pi\left|\mathcal{D}^{\left(1\right)}_{-1}\left(-i\mu_{0}\tau\right)\right|^{2}\\\nonumber&-&14\pi^2\mathrm{Im}\left[D_{-1}\left(i\mu_{0}\tau\right)\mathcal{D}^{\left(1\right)}_{-1}\left(i\overline\mu_0\tau\right)\right]-24\mathrm{Im}\left[\mathcal{D}^{\left(1\right)}_{-1}\left(-i\mu_{0}\tau\right)\mathcal{D}^{\left(2\right)}_{-1}\left(i\overline\mu_0\tau\right)\right]\\&+&8\mathrm{Im}\left[D_{-1}\left(-i\mu_{0}\tau\right)\mathcal{D}^{\left(3\right)}_{-1}\left(i\overline\mu_0\tau\right)\right]+12\pi\mathrm{Re}\left[D_{-1}\left(-i\mu_{0}\tau\right)\mathcal{D}^{\left(2\right)}_{-1}\left(i\overline\mu_0\tau\right)\right]\Big].
	\end{eqnarray} 
	\begin{eqnarray}\label{equ32g}
		\nonumber\mathcal{I}_{5}(\tau)&=&\frac{1}{2^{14}. 5!}\Big[\frac{61\pi^4}{16}\left|D_{-1}\left(-i\mu_{0}\tau\right)\right|^{2}+35\pi^2\left|\mathcal{D}^{\left(1\right)}_{-1}\left(-i\mu_{0}\tau\right)\right|^{2}+30\left|\mathcal{D}^{\left(2\right)}_{-1}\left(-i\mu_{0}\tau\right)\right|^{2}\\\nonumber&+&25\pi^3\mathrm{Im}\left[D_{-1}\left(i\mu_{0}\tau\right)\mathcal{D}^{\left(1\right)}_{-1}\left(i\overline\mu_0\tau\right)\right]+60\pi\mathrm{Im}\left[\mathcal{D}^{\left(1\right)}_{-1}\left(-i\mu_{0}\tau\right)\mathcal{D}^{\left(2\right)}_{-1}\left(i\overline\mu_0\tau\right)\right]\\\nonumber&+&20\pi\mathrm{Im}\left[D_{-1}\left(-i\mu_{0}\tau\right)\mathcal{D}^{\left(3\right)}_{-1}\left(i\overline\mu_0\tau\right)\right]-35\pi^2\mathrm{Re}\left[D_{-1}\left(-i\mu_{0}\tau\right)\mathcal{D}^{\left(2\right)}_{-1}\left(i\overline\mu_0\tau\right)\right]\\&-&40\mathrm{Re}\left[\mathcal{D}^{\left(1\right)}_{-1}\left(-i\mu_{0}\tau\right)\mathcal{D}^{\left(3\right)}_{-1}\left(i\overline\mu_0\tau\right)\right]+10\mathrm{Re}\left[D_{-1}\left(-i\mu_{0}\tau\right)\mathcal{D}^{\left(4\right)}_{-1}\left(i\overline\mu_0\tau\right)\right]\Big].
	\end{eqnarray} 
	Our package, {\bf OscillatoryIntegralAnalytical.wl} returns exactly the same result. Higher order of the integrals can therefore be generated.
	
	\subsection{More results}\label{Sec4.2}
	As seen so far, the traditional LZ problem has led to several hallmarks successes, the incommensurable list of which cannot be enumerated. It has allowed to solve some mathematical problems including the one discussed in this piece notes. We would like to wrap up by listing yet interesting integral relations that are important for level-crossing problems and broadly for other disciplines in Science.
	
	\subsubsection{From double to single integral, $\nu=0$}
\begin{figure}[]
		\vspace{-0.5cm}
		\centering
		\begin{center} 
			\includegraphics[width=8cm, height=6cm]{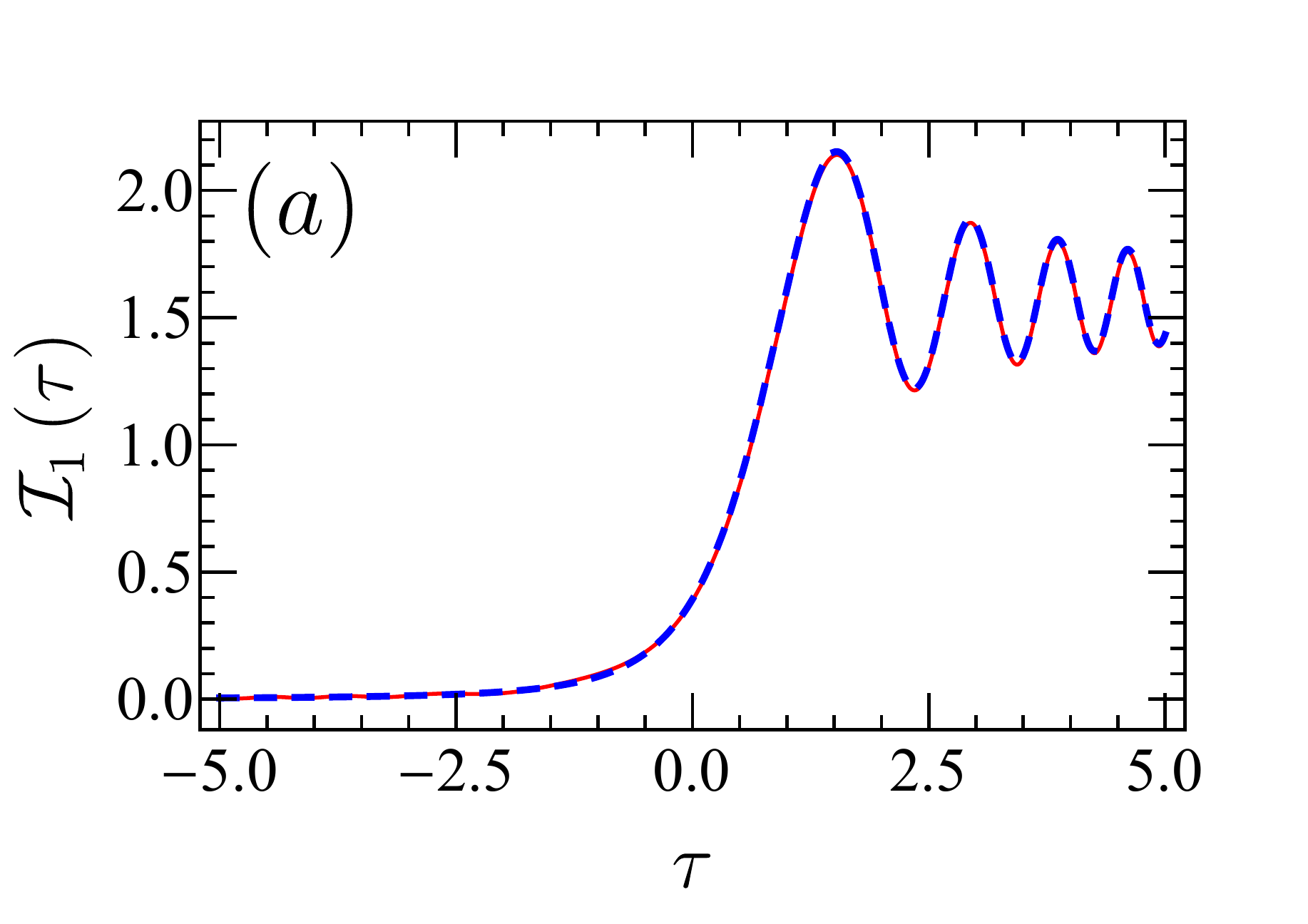}\hspace{-0.75cm}
			\includegraphics[width=8cm, height=6cm]{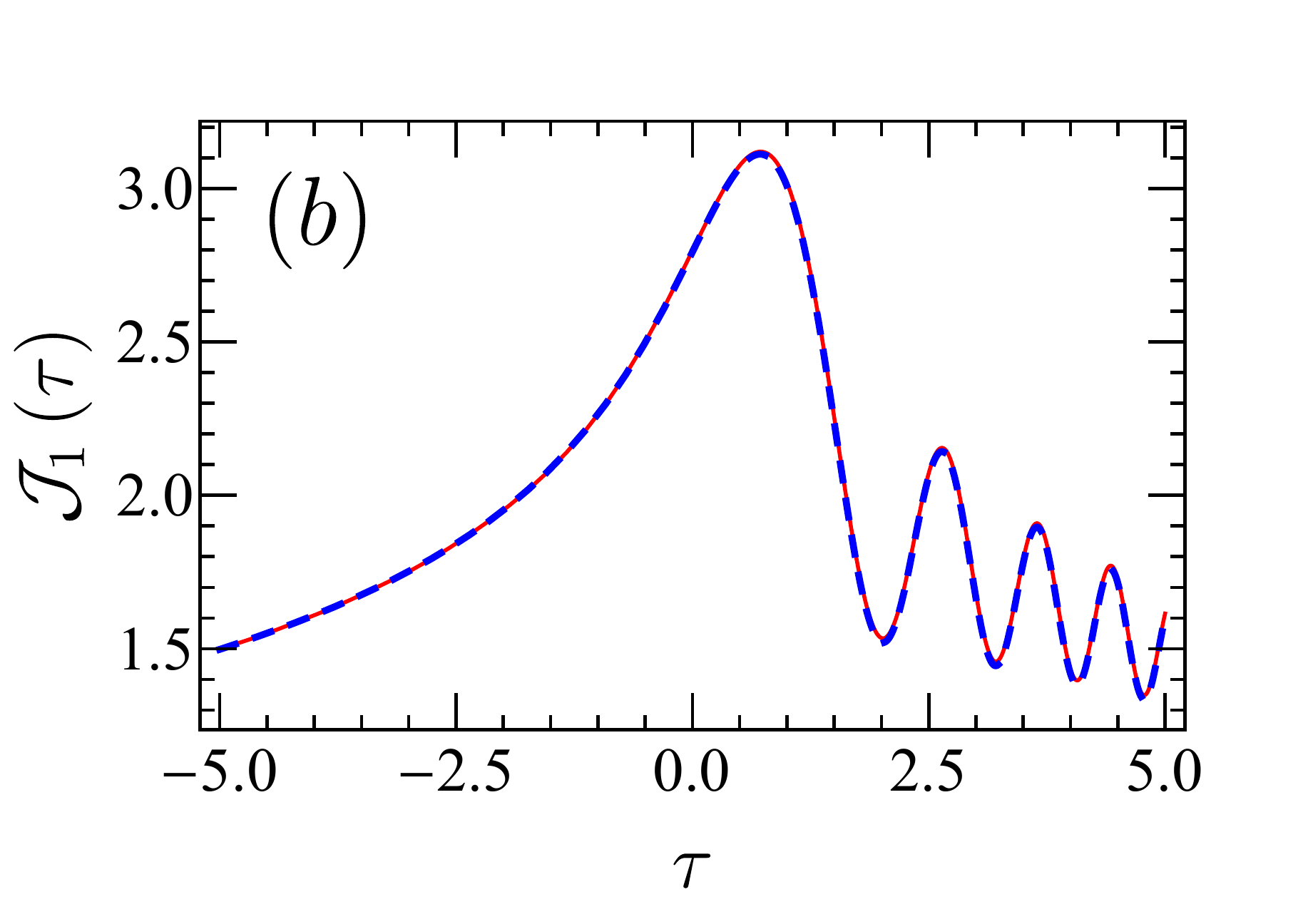}
			\vspace{-0.5cm}
			\caption{Analytical solutions to the integrals $\mathcal{I}_1\left(\tau\right)$ and $\mathcal{J}_1\left(\tau\right)$ versus numerical data.  Solid red lines and blue dashed lines are respectively exact analytical results and numerical data. Simulations are conducted by first transforming the integrals Eqs.\eqref{equ4.12b} and \eqref{equ4.12c} into their equivalent third order ODE. Both data are in excellent quantitative agreement. We observe that $\mathcal{I}_1\left(\tau\right)$ describes oscillations in level populations during non-adiabatic transfer through level-crossing while $\mathcal{J}_1\left(\tau\right)$ describes decay in level population.}
			\label{Figure4}
		\end{center}
	\end{figure}

	The dynamics of the transverse components of the Bloch vectors in \eqref{equ8a} and \eqref{equ8b} processing in the equatorial plane in the unit-radius Bloch sphere are described by integral relations. These are achieved by considering that they are at the origin of the sphere at initial time. Thus, in the PCFs space the phase angles $\varphi_1$ and $\varphi_2$ compensate one another by complex conjugation  and we infer yet interesting results\cite{Kenmoe2013, Brataas2011}
	\begin{eqnarray}\label{equ4.11}
		\nonumber	D_{-i\nu}\left(-i\mu_0\tau\right)D^{*}_{-i\nu-1}\left(-i\mu_0\tau\right)&=&-\mu_0\nu\int_{-\infty}^{\tau}d\tau_1\exp\left(\frac{\mu_0^2}{2}\left[\tau^2-\tau_1^{2}\right]\right)\times\\ &\times&\left[\left|D_{-i\nu-1}\left(-i\mu_0\tau_1\right)\right|^{2}-\frac{1}{\nu}\left|D_{-i\nu}(-i\mu_0\tau_1)\right|^{2}\right], 
	\end{eqnarray} 
	\begin{eqnarray}\label{equ4.12}
		\nonumber	D^{*}_{-i\nu}\left(-i\mu_0\tau\right)D_{-i\nu-1}\left(-i\mu_0\tau\right)&=&-\mu_0^{*}\nu\int_{-\infty}^{\tau}d\tau_1\exp\left(-\frac{\mu_0^2}{2}\left[\tau^2-\tau_1^{2}\right]\right)\times \\&\times&\left[\left|D_{-i\nu-1}\left(-i\mu_0\tau_1\right)\right|^{2}-\frac{1}{\nu}\left|D_{-i\nu}(-i\mu_0\tau_1)\right|^{2}\right]. 
	\end{eqnarray} 
	Here and hereafter in this Subsection $\nu\in{\rm I\!R}$. These integrals are equally numerically verified and we observe a perfect agreement between the data from the left and right hand sides. They play a crucial role in deriving our next results allowing us to fold a double integral into a single integral. Our goal here is to (re)evaluate the following integrals
	\begin{eqnarray}\label{equ4.12b}
		\mathcal{I}_1\left(\tau\right)=\int_{-\infty}^{\tau}d\tau_1\int_{-\infty}^{\tau_1}d\tau_2\cos\left[\tau^2_1-\tau^2_2\right],
	\end{eqnarray} 
	\begin{eqnarray}\label{equ4.12c}
		\mathcal{J}_1\left(\tau\right)=\int_{-\infty}^{\tau}d\tau_1\int_{-\infty}^{\tau_1}d\tau_2\sin\left[\tau^2_1-\tau^2_2\right].
	\end{eqnarray}
	Although $\mathcal{I}_1\left(\tau\right)$ can already be calculated in two different ways, the new strategy adopted here allows us to express it through different functions. 
	Besides the fact that $\mathcal{J}_1\left(\tau\right)$ is less widely known compared to $\mathcal{I}_1\left(\tau\right)$, it remains important for level-crossing problems. While $\mathcal{I}_1\left(\tau\right)$ describes oscillations in level population during non-adiabatic transitions,  $\mathcal{J}_1\left(\tau\right)$ on the other hand describes decay (see Fig.\ref{Figure4}) and deserves to be exactly calculated. It can also be confused with a LZ transition probability in three-level systems.
	
	As already underlined, the integrand in \eqref{equ4.12b} is symmetric under exchange of arguments. For this reason, it can be reduced to the square of two distinct single integrals as $\mathcal{I}_1\left(\tau\right)=\frac{1}{2!}\Big[\int_{-\infty}^{\tau}d\tau_1\cos(\tau_1^2)\Big]^2+\frac{1}{2!}\Big[\int_{-\infty}^{\tau}d\tau_1\sin(\tau_1^2)\Big]^2$.  Such a procedure involved Fresnel integrals\cite{Abramowitz, Fai2024} $C(\tau)=\int d\tau\cos(\tau^2)$ and $S(\tau)=\int d\tau\sin(\tau^2)$. This observation does not hold for \eqref{equ4.12c} as its integrand is anti-symmetric by interchange of arguments. It can therefore not be reduced to single integrals as we did for \eqref{equ4.12b}. Any attempt to direct integration leads to integrals of the forms $\int d\tau\cos(\tau^2)S(\tau)$ and $\int d\tau\sin(\tau^2)S(\tau)$ which to the best of our knowledge cannot yet be exactly evaluated using actual techniques in both Mathematics and Physics. In order to resolve this issue, we go back to \eqref{equ4.11} and \eqref{equ4.12}, set $\nu=0$ and take separately the sum and difference of the resulting equations. Thereby,
	\begin{eqnarray}\label{equ4.12bb}
		\mathcal{I}_1\left(\tau\right)= -\frac{i}{2\sqrt{2}}\int_{-\infty}^{\tau}d\tau_1\left(e^{i\pi/4}e^{i\tau^2_1/2}D_{-1}\left(-i\mu_0\tau\right)-e^{-i\pi/4}e^{-i\tau^2_1/2}D^{*}_{-1}\left(-i\mu_0\tau\right)\right),
	\end{eqnarray} 
	\begin{eqnarray}\label{equ4.12cc}
		\mathcal{J}_1\left(\tau\right)= -\frac{1}{2\sqrt{2}}\int_{-\infty}^{\tau}d\tau_1\left(e^{i\pi/4}e^{i\tau^2_1/2}D_{-1}\left(-i\mu_0\tau\right)+e^{-i\pi/4}e^{-i\tau^2_1/2}D^{*}_{-1}\left(-i\mu_0\tau\right)\right).
	\end{eqnarray} 
	We have thus transformed double integrals with highly oscillatory integrands into single integrals with less oscillatory kernels.  It should be observed that the two integrands in the right hand sides are mutually complex conjugates. Then, knowing one of the integrals is enough to completely evaluate $\mathcal{I}_1\left(\tau\right)$ and $\mathcal{J}_1\left(\tau\right)$. This stems from the following anti-derivative,
	\begin{eqnarray}\label{equ4.12d}
		\int d\tau e^{i\tau^2/2}D_{-1}\left(-i\mu_0\tau\right)=i\mu_0\mathcal{R}\left(\tau\right),
	\end{eqnarray} 
	where 
	\begin{eqnarray}\label{equ4.12e}
		\mathcal{R}\left(\tau\right) = \frac{\pi}{4}{\rm erf}\left(\frac{\mu_0\tau}{\sqrt{2}}\right)+\frac{\tau^2}{2} {}_{2}F_{2}\left(\frac{3}{2},2,i\tau^2\right).
	\end{eqnarray} 
	Here, ${}_{2}F_{2}\left(a_1,a_2;b_1,b_2; z\right)$ is a generalized Hypergeometric function\cite{Abramowitz, Fai2024}. Because $a_1=a_2=1$, we have used the shorthand notation in \eqref{equ4.12e}. Returning to \eqref{equ4.12b} and \eqref{equ4.12c} equipped with \eqref{equ4.12d}, we achieve,
	\begin{eqnarray}\label{equ4.12ee}
		\mathcal{I}_1\left(\tau\right)  = {\rm Re}\left[\mathcal{R}\left(\tau\right)-\mathcal{R}\left(-\infty\right)\right],
	\end{eqnarray} 
	\begin{eqnarray}\label{equ4.12f}
		\mathcal{J}_1\left(\tau\right)  = -{\rm Im}\left[\mathcal{R}\left(\tau\right)-\mathcal{R}\left(-\infty\right)\right].
	\end{eqnarray} 
	These results are very important for the aforementioned reasons. They establish yet connections between PCFs and Hypergeometric functions.  Further insights into level-crossing problems can then be gained from here. Both integrals describe accumulation of oscillatory phase through virtual interactions. We can assert that the kernel in $\mathcal{I}_1\left(\tau\right)$ being an even function,  has a symmetric structure leading to constructive accumulation of virtual interactions while the odd kernel in $\mathcal{J}_1\left(\tau\right)$  with an anti-symmetric structure causes destructive cancellation. Indeed, let us compute
	\begin{eqnarray}\label{equ4.12g}
		\mathcal{I}_1\left(\tau\right) \pm i\mathcal{J}_1\left(\tau\right)  = \int_{-\infty}^{\tau}d\tau_1\int_{-\infty}^{\tau_1}d\tau_2\exp\left(\pm i\left[\tau^2_1-\tau^2_2\right]\right).
	\end{eqnarray} 
	We learn from Fig.\ref{Figure4} that the real part of \eqref{equ4.12g} describes population transfer while the imaginary part describes population loss. In the limit $\tau\to\infty$ the double integral in \eqref{equ4.12g} reduces to $\mathcal{I}_1\left(\infty\right)$ as $\mathcal{J}_1\left(\infty\right)=0$.
	
	
	As proned so far, the only mean to verify our analytical calculations is through numerical simulations. Because $\mathcal{J}_{1}(\tau)$ is also a highly oscillating double integral, its numerical integration is found to be extremely slow, unstable with low precision i.e. merely not reliable. On the other hand, transforming it into an equivalent third order ODE makes it much faster, stable with very high precision (up to 30 digits). Thus,
	\begin{eqnarray}\label{equ4.25}
		\frac{d^3}{d\tau^3}\mathcal{J}_{1}(\tau)-\frac{1}{\tau}\frac{d^2}{d\tau^2}\mathcal{J}_{1}(\tau)+4\tau^2\frac{d}{d\tau}\mathcal{J}_{1}(\tau)=-2\tau,
	\end{eqnarray}
	solved with initial conditions
	\begin{eqnarray}\label{equ4.26}
		\frac{d^2}{d\tau^2}\mathcal{J}_{1}(\tau)|_{\tau\to-\infty}= \frac{d}{d\tau}\mathcal{J}_{1}(\tau)|_{\tau\to-\infty}=\mathcal{J}_{1}(-\infty)=0.
	\end{eqnarray}
	Numerical results associated with these equations are displayed in Fig.\ref{Figure4}.

	\subsubsection{From double to single integral, arbitrary $\nu$}
	For $\nu=0$, we have been able to derive a closed-form solution to yet another important highly oscillating double integral. This was achieved by exploiting again its connection with the celebrated  LZ problem and has helped us reducing a double integral of complicated functions to a single integral of simple functions that we were ultimately able to integrate. We now want to extend this philosophy to arbitrary $\nu$. Exactly as we did above, we achieve
	\begin{eqnarray}\label{equ4.30}
		\nonumber \int_{-\infty}^{\tau}d\tau_1\left(e^{-i\pi/4}D_{-i\nu}\left(-i\mu_0\tau_1\right)D^{*}_{-i\nu-1}\left(-i\mu_0\tau_1\right)-e^{i\pi/4}D^{*}_{-i\nu}\left(-i\mu_0\tau_1\right)D_{-i\nu-1}\left(-i\mu_0\tau_1\right)\right)\\\nonumber=2\nu\sqrt{2}\int_{-\infty}^{\tau}d\tau_1\int_{-\infty}^{\tau_1}d\tau_2\cos\left(\tau_1^2-\tau_2^{2}\right) \left[\left|D_{-i\nu-1}\left(-i\mu_0\tau_2\right)\right|^{2}-\frac{1}{\nu}\left|D_{-i\nu}(-i\mu_0\tau_2)\right|^{2}\right], \quad \nu\in{\rm I\!R},\\
	\end{eqnarray} 
	\begin{eqnarray}\label{equ4.31}
		\nonumber \int_{-\infty}^{\tau}d\tau_1\left(e^{-i\pi/4}D_{-i\nu}\left(-i\mu_0\tau_1\right)D^{*}_{-i\nu-1}\left(-i\mu_0\tau_1\right)+e^{i\pi/4}D^{*}_{-i\nu}\left(-i\mu_0\tau_1\right)D_{-i\nu-1}\left(-i\mu_0\tau_1\right)\right)\\\nonumber=2\nu\sqrt{2}\int_{-\infty}^{\tau}d\tau_1\int_{-\infty}^{\tau_1}d\tau_2\sin\left(\tau_1^2-\tau_2^{2}\right) \left[\left|D_{-i\nu-1}\left(-i\mu_0\tau_2\right)\right|^{2}-\frac{1}{\nu}\left|D_{-i\nu}(-i\mu_0\tau_2)\right|^{2}\right], \quad \nu\in{\rm I\!R}.\\
	\end{eqnarray} 
	These relations are also verified numerically. Eq.\eqref{equ4.30} is reminiscent of \eqref{equ7d} in the sense that knowing a double integral could help deriving the following quadruple integral. Note that the integrand in the left-hand side of \eqref{equ4.30} is nothing but $u_y(\tau)$ while that of \eqref{equ4.31} is $u_x(\tau)$. Going back to our description of the LZ problem in the Bloch picture, we find the trivial connection $\dot{u}_z(\tau)=2\sqrt{2\nu}u_y(\tau)$ whereby
	\begin{eqnarray}\label{equ4.33}
		\nonumber	\hspace{-1.7cm}\left|D_{-i\nu-1}\left(-i\mu\tau\right)\right|^{2}&=&i\sqrt{2}\int_{-\infty}^{\tau}d\tau_1\Big[e^{-i\pi/4}D_{-i\nu}\left(-i\mu\tau_1\right)D^{*}_{-i\nu-1}\left(-i\mu\tau_1\right)\\&-&e^{i\pi/4}D^{*}_{-i\nu}\left(-i\mu\tau_1\right)D_{-i\nu-1}\left(-i\mu\tau_1\right)\Big], \quad \nu\in{\rm I\!R}. 
	\end{eqnarray} 
	We cannot establish such a simple relation between $u_x(\tau)$ and $u_z(\tau)$. As a consequence, the right-hand side of \eqref{equ4.31} cannot be evaluated at this stage and is left for future projects. Moreover, combining \eqref{equ4.33} and \eqref{equ4.30} leads us to
	\begin{eqnarray}\label{equ4.14}
	\nonumber	\left|D_{-i\nu-1}\left(-i\mu\tau\right)\right|^{2}&=&-4\nu\int_{-\infty}^{\tau}d\tau_1\int_{-\infty}^{\tau_1}d\tau_2\cos\left(\tau_1^2-\tau_2^{2}\right)\times\\&\times& \Big[\left|D_{-i\nu-1}\left(-i\mu\tau_2\right)\right|^{2}-\frac{1}{\nu}\left|D_{-i\nu}(-i\mu\tau_2)\right|^{2}\Big], \quad \nu\in{\rm I\!R}.
	\end{eqnarray} 
	Setting $\nu=0$ in this relation while having in mind $\left|D_{0}(-i\mu\tau)\right|^{2}=1$ yields $\mathcal{I}_1\left(\tau\right)$ without further calculations. Taking the derivative of \eqref{equ4.14} with respect to $\nu$, further setting $\nu=0$ and considering \eqref{equ32b} gives
	\begin{eqnarray}\label{equ4.31a}
		\nonumber\pi\left|D_{-1}\left(-i\mu_{0}\tau\right)\right|^{2}+4\mathrm{Im}\left[D_{-1}\left(-i\mu_{0}\tau\right)\mathcal{D}^{\left(1\right)}_{-1}\left(i\overline\mu_0\tau\right)\right]&=&16\int^\tau_{-\infty}d\tau_{1}\int^{\tau_{1}}_{-\infty}d\tau_{2}\cos[\tau^{2}_{1}-\tau^{2}_{2}]\times\\&\times&{\left|D_{-1}\left(-i\mu_0\tau_2\right)\right|}^{2},
	\end{eqnarray}  
	which is nothing but $\mathcal{I}_2\left(\tau\right)$. Direct calculation of $\mathcal{I}_1\left(\tau\right)$ establishes a relation between Fresnel integrals and PCFs as (see also Ref.\onlinecite{Kiselev2013, Kovaleva, Garanin2005, Yan})
	\begin{eqnarray}\label{equ4.15}
		\left|D_{-1}\left(-i\mu_0\tau\right)\right|^{2}=\pi\left({\left[\frac{1}{2}+C\left(\sqrt{\frac{2}{\pi}}\tau\right)\right]}^{2}+{\left[\frac{1}{2}+S\left(\sqrt{\frac{2}{\pi}}\tau\right)\right]}^{2}\right).
	\end{eqnarray} 
	By virtue of the fact that $C\left(\infty\right)=S\left(\infty\right)=1/2$, we also verify that $\left|D_{-1}\left(-i\mu_0\tau\right)\right|=\sqrt{2\pi}$ when $\tau\to\infty$ as indicated earlier. Similarly, as $C\left(0\right)=S\left(0\right)=0$, then $\left|D_{-1}\left(-i\mu_0\tau\right)\right|=\sqrt{\pi/2}$ for $\tau\to0$. At the opposite far remote time $C\left(-\infty\right)=S\left(-\infty\right)=-1/2$ hence $\left|D_{-1}\left(-i\mu_0\tau\right)\right|=0$ as $\tau\to -\infty$.
	In the same vein, initial condition and the conservation of population lead to important relations between PCFs,
	\begin{eqnarray}\label{equ4.16}
		\lim_{\tau\to-\infty}\left(-\nu e^{-\pi\nu/2}\left[\left|D_{-i\nu-1}\left(-i\mu_0\tau\right)\right|^{2}-\frac{1}{\nu}\left|D_{-i\nu}(-i\mu_0\tau)\right|^{2}\right]\right)=1,
	\end{eqnarray} 
	\begin{eqnarray}\label{equ4.17}
		\nu e^{-\pi\nu/2}\left[\left|D_{-i\nu-1}\left(-i\mu_0\tau\right)\right|^{2}+\frac{1}{\nu}\left|D_{-i\nu}(-i\mu_0\tau)\right|^{2}\right]=1.
	\end{eqnarray}  
	
	\section{Concluding Remarks}\label{Sec6}
	We derived a closed-form exact solution to a highly nontrivial multidimensional oscillatory integral. Except at $\tau=\infty$, this was up to now inaccessible due to complications inherent to the complex derivative of Parabolic Cylinder functions with respect to the index. Calculations are carried out by relating the integral to one of the rare standard problems in non-stationary quantum Mechanics that admits an exact solution: the two-level Landau-Zener problem\cite{Landau1932, Zener1932, Stuckelberg1932, Majorana1932}. Although the two problems have simple connections at $\tau=\infty$ and complicated ones at finite time, we were able to circumvent this difficulty  by introducing modified PCFs. This allowed us to perform the desired derivative and achieve our goal. The present calculations open a pathway toward generalizing such methods to more complex time-dependent systems in non-stationary quantum mechanics. Our results have great potentials for improvement in reexamining several problems in level-crossing Physics. In Ref.\onlinecite{Kenmoe2013} for example, was discussed the effects of classical slow noise on LZ transitions at finite-time. Only approximated solutions valid in the sudden transitions limit were proposed as the problem involved $\mathcal{I}_k$. Because only $\mathcal{I}_1$ could be evaluated, the remaining corrections were exponentiated allowing an easy averaging over the Gaussian distribution of the noise. In Ref.\onlinecite{Kiselev2013} was proposed a general formulation of three-level LZ problem. The resulting equations for the density matrix involving $\mathcal{I}_k(\tau)$ could only be solved perturbatively up to the first order. Now, thanks to the present results, higher order corrections can be included and might help gaining further insights into the aforementioned problems. Solving these problems lies beyond the scoop of the present paper.
	
	
	\appendix
	\section{Bell Polynomials}
	We want in this Appendix to present a few explicit forms of Bell Polynomials. In our case, we only needed the complete  $\mathcal{B}_{n}$ and incomplete  $\mathcal{B}_{n,j}$ Bell. They are connected through,
	\begin{eqnarray}\label{equA1}
		\mathcal{B}_{n}\left(\phi^{\left(0\right)},\phi^{\left(1\right)},\cdots,\phi^{\left(n-2\right)},\phi^{\left(n-1\right)}\right)=\sum_{j=1}^{n}\mathcal{B}_{n,j}\left(\phi^{\left(0\right)},\phi^{\left(1\right)},\cdots,\phi^{\left(n-2\right)},\phi^{\left(n-1\right)}\right),
	\end{eqnarray}
	Hereby, 
	\begin{eqnarray}\label{equA2}
		\mathcal{B}_{1}\left(\phi^{\left(0\right)}\right)=\phi^{\left(0\right)},
	\end{eqnarray}
	\begin{eqnarray}\label{equA3}
		\mathcal{B}_{2}\left(\phi^{\left(0\right)},\phi^{\left(1\right)}\right)=\left(\phi^{\left(0\right)}\right)^2+\phi^{\left(1\right)},
	\end{eqnarray}
	\begin{eqnarray}\label{equA4}
		\mathcal{B}_{3}\left(\phi^{\left(0\right)},\phi^{\left(1\right)},\phi^{\left(2\right)}\right)=\left(\phi^{\left(0\right)}\right)^3+3\phi^{\left(0\right)}\phi^{\left(1\right)}+\phi^{\left(2\right)},
	\end{eqnarray}
	\begin{eqnarray}\label{equA5}
		\mathcal{B}_{4}\left(\phi^{\left(0\right)},\phi^{\left(1\right)},\phi^{\left(2\right)},\phi^{\left(3\right)}\right)=\left(\phi^{\left(0\right)}\right)^4+6\left(\phi^{\left(0\right)}\right)^2\phi^{\left(1\right)}+3\left(\phi^{\left(1\right)}\right)^2+
		4\phi^{\left(0\right)}\phi^{\left(2\right)}+\phi^{\left(3\right)},
	\end{eqnarray}
	\begin{eqnarray}\label{equA6}
		\nonumber	\mathcal{B}_{5}\left(\phi^{\left(0\right)},\phi^{\left(1\right)},\phi^{\left(2\right)},\phi^{\left(3\right)},\phi^{\left(4\right)}\right)=\left(\phi^{\left(0\right)}\right)^5+10\left(\phi^{\left(0\right)}\right)^3\phi^{\left(1\right)}+15\phi^{\left(0\right)}\left(\phi^{\left(1\right)}\right)^2+
		10\left(\phi^{\left(0\right)}\right)^2\phi^{\left(2\right)}\\+
		10\phi^{\left(1\right)}\phi^{\left(2\right)}+5\phi^{\left(0\right)}\phi^{\left(3\right)}+\phi^{\left(4\right)}.
	\end{eqnarray}
	
	\section{Acknowledgments}
	This work is part of the project "Building a Comprehensive Quantum Theory for Many-Body Localization in Spin chains". MBK has received financial support from the Alexander von Humboldt foundation through the scheme of Georg-Forster fellowship for Postdocs. He is also very grateful to Prof. John Schliemann for the very warm hospitality at the University of Regensburg where a part of this paper was written.

	\bibliography{Mybib}

\begin{thebibliography}{54}%
\makeatletter
\providecommand \@ifxundefined [1]{%
 \@ifx{#1\undefined}
}%
\providecommand \@ifnum [1]{%
 \ifnum #1\expandafter \@firstoftwo
 \else \expandafter \@secondoftwo
 \fi
}%
\providecommand \@ifx [1]{%
 \ifx #1\expandafter \@firstoftwo
 \else \expandafter \@secondoftwo
 \fi
}%
\providecommand \natexlab [1]{#1}%
\providecommand \enquote  [1]{``#1''}%
\providecommand \bibnamefont  [1]{#1}%
\providecommand \bibfnamefont [1]{#1}%
\providecommand \citenamefont [1]{#1}%
\providecommand \href@noop [0]{\@secondoftwo}%
\providecommand \href [0]{\begingroup \@sanitize@url \@href}%
\providecommand \@href[1]{\@@startlink{#1}\@@href}%
\providecommand \@@href[1]{\endgroup#1\@@endlink}%
\providecommand \@sanitize@url [0]{\catcode `\\12\catcode `\$12\catcode
  `\&12\catcode `\#12\catcode `\^12\catcode `\_12\catcode `\%12\relax}%
\providecommand \@@startlink[1]{}%
\providecommand \@@endlink[0]{}%
\providecommand \url  [0]{\begingroup\@sanitize@url \@url }%
\providecommand \@url [1]{\endgroup\@href {#1}{\urlprefix }}%
\providecommand \urlprefix  [0]{URL }%
\providecommand \Eprint [0]{\href }%
\providecommand \doibase [0]{https://doi.org/}%
\providecommand \selectlanguage [0]{\@gobble}%
\providecommand \bibinfo  [0]{\@secondoftwo}%
\providecommand \bibfield  [0]{\@secondoftwo}%
\providecommand \translation [1]{[#1]}%
\providecommand \BibitemOpen [0]{}%
\providecommand \bibitemStop [0]{}%
\providecommand \bibitemNoStop [0]{.\EOS\space}%
\providecommand \EOS [0]{\spacefactor3000\relax}%
\providecommand \BibitemShut  [1]{\csname bibitem#1\endcsname}%
\let\auto@bib@innerbib\@empty
\bibitem [{\citenamefont {Kayanuma}(1984)}]{Kayanuma1984}%
  \BibitemOpen
  \bibfield  {author} {\bibinfo {author} {\bibfnamefont {Y.}~\bibnamefont
  {Kayanuma}},\ }\bibfield  {title} {\enquote {\bibinfo {title} {Nonadiabatic
  transitions in level crossing with energy fluctuation. {I}. {A}nalytical
  {I}nvestigations},}\ }\href {https://doi.org/10.1143/JPSJ.53.108} {\bibfield
  {journal} {\bibinfo  {journal} {Journal of the Physical Society of Japan}\
  }\textbf {\bibinfo {volume} {53}},\ \bibinfo {pages} {108--117} (\bibinfo
  {year} {1984})},\ \Eprint
  {https://arxiv.org/abs/https://doi.org/10.1143/JPSJ.53.108}
  {https://doi.org/10.1143/JPSJ.53.108} \BibitemShut {NoStop}%
\bibitem [{\citenamefont {Kayanuma}\ and\ \citenamefont
  {Fukuchi}(1985)}]{Kayanuma1985}%
  \BibitemOpen
  \bibfield  {author} {\bibinfo {author} {\bibfnamefont {Y.}~\bibnamefont
  {Kayanuma}}\ and\ \bibinfo {author} {\bibfnamefont {S.}~\bibnamefont
  {Fukuchi}},\ }\bibfield  {title} {\enquote {\bibinfo {title} {On the
  probability of non-adiabatic transitions in multiple level crossings},}\
  }\href {https://doi.org/10.1088/0022-3700/18/20/012} {\bibfield  {journal}
  {\bibinfo  {journal} {Journal of Physics B: Atomic and Molecular Physics}\
  }\textbf {\bibinfo {volume} {18}},\ \bibinfo {pages} {4089} (\bibinfo {year}
  {1985})}\BibitemShut {NoStop}%
\bibitem [{\citenamefont {Kayanuma}\ and\ \citenamefont
  {Nakayama}(1998)}]{Kayanuma1998}%
  \BibitemOpen
  \bibfield  {author} {\bibinfo {author} {\bibfnamefont {Y.}~\bibnamefont
  {Kayanuma}}\ and\ \bibinfo {author} {\bibfnamefont {H.}~\bibnamefont
  {Nakayama}},\ }\bibfield  {title} {\enquote {\bibinfo {title} {Nonadiabatic
  transition at a level crossing with dissipation},}\ }\href
  {https://doi.org/10.1103/PhysRevB.57.13099} {\bibfield  {journal} {\bibinfo
  {journal} {Phys. Rev. B}\ }\textbf {\bibinfo {volume} {57}},\ \bibinfo
  {pages} {13099--13112} (\bibinfo {year} {1998})}\BibitemShut {NoStop}%
\bibitem [{\citenamefont {Demkov}\ and\ \citenamefont
  {Ostrovsky}(2001)}]{Demkov2001}%
  \BibitemOpen
  \bibfield  {author} {\bibinfo {author} {\bibfnamefont {Y.~N.}\ \bibnamefont
  {Demkov}}\ and\ \bibinfo {author} {\bibfnamefont {V.~N.}\ \bibnamefont
  {Ostrovsky}},\ }\bibfield  {title} {\enquote {\bibinfo {title} {The exact
  solution of the multistate {L}andau-{Z}ener type model: the generalized
  bow-tie model},}\ }\href {https://doi.org/10.1088/0953-4075/34/12/309}
  {\bibfield  {journal} {\bibinfo  {journal} {Journal of Physics B: Atomic,
  Molecular and Optical Physics}\ }\textbf {\bibinfo {volume} {34}},\ \bibinfo
  {pages} {2419} (\bibinfo {year} {2001})}\BibitemShut {NoStop}%
\bibitem [{\citenamefont {Volkov}\ and\ \citenamefont
  {Ostrovsky}(2004)}]{Volkov2004}%
  \BibitemOpen
  \bibfield  {author} {\bibinfo {author} {\bibfnamefont {M.~V.}\ \bibnamefont
  {Volkov}}\ and\ \bibinfo {author} {\bibfnamefont {V.~N.}\ \bibnamefont
  {Ostrovsky}},\ }\bibfield  {title} {\enquote {\bibinfo {title} {Exact results
  for survival probability in the multistate {L}andau-{Z}ener model},}\ }\href
  {https://doi.org/10.1088/0953-4075/37/20/003} {\bibfield  {journal} {\bibinfo
   {journal} {Journal of Physics B: Atomic, Molecular and Optical Physics}\
  }\textbf {\bibinfo {volume} {37}},\ \bibinfo {pages} {4069} (\bibinfo {year}
  {2004})}\BibitemShut {NoStop}%
\bibitem [{\citenamefont {Saito}\ \emph {et~al.}(2007)\citenamefont {Saito},
  \citenamefont {Wubs}, \citenamefont {Kohler}, \citenamefont {Kayanuma},\ and\
  \citenamefont {H\"anggi}}]{Saito2007}%
  \BibitemOpen
  \bibfield  {author} {\bibinfo {author} {\bibfnamefont {K.}~\bibnamefont
  {Saito}}, \bibinfo {author} {\bibfnamefont {M.}~\bibnamefont {Wubs}},
  \bibinfo {author} {\bibfnamefont {S.}~\bibnamefont {Kohler}}, \bibinfo
  {author} {\bibfnamefont {Y.}~\bibnamefont {Kayanuma}},\ and\ \bibinfo
  {author} {\bibfnamefont {P.}~\bibnamefont {H\"anggi}},\ }\bibfield  {title}
  {\enquote {\bibinfo {title} {Dissipative {L}andau-{Z}ener transitions of a
  qubit: Bath-specific and universal behavior},}\ }\href
  {https://doi.org/10.1103/PhysRevB.75.214308} {\bibfield  {journal} {\bibinfo
  {journal} {Phys. Rev. B}\ }\textbf {\bibinfo {volume} {75}},\ \bibinfo
  {pages} {214308} (\bibinfo {year} {2007})}\BibitemShut {NoStop}%
\bibitem [{\citenamefont {Rojo}(2010)}]{Rojo2010}%
  \BibitemOpen
  \bibfield  {author} {\bibinfo {author} {\bibfnamefont {A.~G.}\ \bibnamefont
  {Rojo}},\ }\bibfield  {title} {\enquote {\bibinfo {title} {Matrix exponential
  solution of the {L}andau-{Z}ener problem},}\ }\href
  {https://doi.org/https://doi.org/10.48550/arXiv.1004.2914} {\bibfield
  {journal} {\bibinfo  {journal} {ArXiv}\ } (\bibinfo {year} {2010}),\
  https://doi.org/10.48550/arXiv.1004.2914}\BibitemShut {NoStop}%
\bibitem [{\citenamefont {Glasbrenner}\ and\ \citenamefont
  {Schleich}(2023)}]{Glasbrenner2023}%
  \BibitemOpen
  \bibfield  {author} {\bibinfo {author} {\bibfnamefont {E.~P.}\ \bibnamefont
  {Glasbrenner}}\ and\ \bibinfo {author} {\bibfnamefont {W.~P.}\ \bibnamefont
  {Schleich}},\ }\bibfield  {title} {\enquote {\bibinfo {title} {The
  {L}andau–{Z}ener formula made simple},}\ }\href
  {https://doi.org/10.1088/1361-6455/acc774} {\bibfield  {journal} {\bibinfo
  {journal} {Journal of Physics B: Atomic, Molecular and Optical Physics}\
  }\textbf {\bibinfo {volume} {56}},\ \bibinfo {pages} {104001} (\bibinfo
  {year} {2023})}\BibitemShut {NoStop}%
\bibitem [{\citenamefont {Shevchenko}, \citenamefont {Ashhab},\ and\
  \citenamefont {Nori}(2010)}]{Shevchenko2010}%
  \BibitemOpen
  \bibfield  {author} {\bibinfo {author} {\bibfnamefont {S.}~\bibnamefont
  {Shevchenko}}, \bibinfo {author} {\bibfnamefont {S.}~\bibnamefont {Ashhab}},\
  and\ \bibinfo {author} {\bibfnamefont {F.}~\bibnamefont {Nori}},\ }\bibfield
  {title} {\enquote {\bibinfo {title} {{L}andau-{Z}ener–stückelberg
  interferometry},}\ }\href
  {https://doi.org/https://doi.org/10.1016/j.physrep.2010.03.002} {\bibfield
  {journal} {\bibinfo  {journal} {Physics Reports}\ }\textbf {\bibinfo {volume}
  {492}},\ \bibinfo {pages} {1--30} (\bibinfo {year} {2010})}\BibitemShut
  {NoStop}%
\bibitem [{\citenamefont {Kiselev}, \citenamefont {Kikoin},\ and\ \citenamefont
  {Kenmoe}(2013)}]{Kiselev2013}%
  \BibitemOpen
  \bibfield  {author} {\bibinfo {author} {\bibfnamefont {M.~N.}\ \bibnamefont
  {Kiselev}}, \bibinfo {author} {\bibfnamefont {K.}~\bibnamefont {Kikoin}},\
  and\ \bibinfo {author} {\bibfnamefont {M.~B.}\ \bibnamefont {Kenmoe}},\
  }\bibfield  {title} {\enquote {\bibinfo {title} {{SU}(3) {L}andau-{Z}ener
  interferometry},}\ }\href {https://doi.org/10.1209/0295-5075/104/57004}
  {\bibfield  {journal} {\bibinfo  {journal} {Europhysics Letters}\ }\textbf
  {\bibinfo {volume} {104}},\ \bibinfo {pages} {57004} (\bibinfo {year}
  {2013})}\BibitemShut {NoStop}%
\bibitem [{\citenamefont {Kenmoe}, \citenamefont {Tchapda},\ and\ \citenamefont
  {Fai}(2017)}]{Kenmoe2017}%
  \BibitemOpen
  \bibfield  {author} {\bibinfo {author} {\bibfnamefont {M.~B.}\ \bibnamefont
  {Kenmoe}}, \bibinfo {author} {\bibfnamefont {A.~B.}\ \bibnamefont
  {Tchapda}},\ and\ \bibinfo {author} {\bibfnamefont {L.~C.}\ \bibnamefont
  {Fai}},\ }\bibfield  {title} {\enquote {\bibinfo {title} {{SU}(3)
  {L}andau-{Z}ener interferometry with a transverse periodic drive},}\ }\href
  {https://doi.org/10.1103/PhysRevB.96.125126} {\bibfield  {journal} {\bibinfo
  {journal} {Phys. Rev. B}\ }\textbf {\bibinfo {volume} {96}},\ \bibinfo
  {pages} {125126} (\bibinfo {year} {2017})}\BibitemShut {NoStop}%
\bibitem [{\citenamefont {Feldbrugge}\ and\ \citenamefont
  {Jones}(2025)}]{Job2025}%
  \BibitemOpen
  \bibfield  {author} {\bibinfo {author} {\bibfnamefont {J.}~\bibnamefont
  {Feldbrugge}}\ and\ \bibinfo {author} {\bibfnamefont {J.~Y.~L.}\ \bibnamefont
  {Jones}},\ }\bibfield  {title} {\enquote {\bibinfo {title} {Efficient
  evaluation of real-time path integrals},}\ }\href
  {https://doi.org/10.1103/PhysRevD.111.083524} {\bibfield  {journal} {\bibinfo
   {journal} {Phys. Rev. D}\ }\textbf {\bibinfo {volume} {111}},\ \bibinfo
  {pages} {083524} (\bibinfo {year} {2025})}\BibitemShut {NoStop}%
\bibitem [{\citenamefont {Hillion}(1997)}]{Diffraction}%
  \BibitemOpen
  \bibfield  {author} {\bibinfo {author} {\bibfnamefont {P.}~\bibnamefont
  {Hillion}},\ }\bibfield  {title} {\enquote {\bibinfo {title} {Diffraction and
  weber functions},}\ }\href {http://www.jstor.org/stable/2951932} {\bibfield
  {journal} {\bibinfo  {journal} {SIAM Journal on Applied Mathematics}\
  }\textbf {\bibinfo {volume} {57}},\ \bibinfo {pages} {1702--1715} (\bibinfo
  {year} {1997})}\BibitemShut {NoStop}%
\bibitem [{\citenamefont {Moshinsky}(1952)}]{Moshinsky}%
  \BibitemOpen
  \bibfield  {author} {\bibinfo {author} {\bibfnamefont {M.}~\bibnamefont
  {Moshinsky}},\ }\bibfield  {title} {\enquote {\bibinfo {title} {Diffraction
  in time},}\ }\href {https://doi.org/10.1103/PhysRev.88.625} {\bibfield
  {journal} {\bibinfo  {journal} {Phys. Rev.}\ }\textbf {\bibinfo {volume}
  {88}},\ \bibinfo {pages} {625--631} (\bibinfo {year} {1952})}\BibitemShut
  {NoStop}%
\bibitem [{\citenamefont {Sornette}(2014)}]{Sornette2014}%
  \BibitemOpen
  \bibfield  {author} {\bibinfo {author} {\bibfnamefont {D.}~\bibnamefont
  {Sornette}},\ }\bibfield  {title} {\enquote {\bibinfo {title} {Physics and
  financial economics (1776–2014): puzzles, {I}sing and agent-based
  models},}\ }\href {https://doi.org/10.1088/0034-4885/77/6/062001} {\bibfield
  {journal} {\bibinfo  {journal} {Reports on Progress in Physics}\ }\textbf
  {\bibinfo {volume} {77}},\ \bibinfo {pages} {062001} (\bibinfo {year}
  {2014})}\BibitemShut {NoStop}%
\bibitem [{\citenamefont {Wang}\ \emph {et~al.}(2023)\citenamefont {Wang},
  \citenamefont {Qin}, \citenamefont {Zhao}, \citenamefont {Ye}, \citenamefont
  {Longhi}, \citenamefont {Lu},\ and\ \citenamefont {Wang}}]{Shulin2023}%
  \BibitemOpen
  \bibfield  {author} {\bibinfo {author} {\bibfnamefont {S.}~\bibnamefont
  {Wang}}, \bibinfo {author} {\bibfnamefont {C.}~\bibnamefont {Qin}}, \bibinfo
  {author} {\bibfnamefont {L.}~\bibnamefont {Zhao}}, \bibinfo {author}
  {\bibfnamefont {H.}~\bibnamefont {Ye}}, \bibinfo {author} {\bibfnamefont
  {S.}~\bibnamefont {Longhi}}, \bibinfo {author} {\bibfnamefont
  {P.}~\bibnamefont {Lu}},\ and\ \bibinfo {author} {\bibfnamefont
  {B.}~\bibnamefont {Wang}},\ }\bibfield  {title} {\enquote {\bibinfo {title}
  {Photonic {F}loquet {L}andau-{Z}ener tunneling and temporal beam
  splitters},}\ }\href {https://doi.org/10.1126/sciadv.adh0415} {\bibfield
  {journal} {\bibinfo  {journal} {Science Advances}\ }\textbf {\bibinfo
  {volume} {9}},\ \bibinfo {pages} {eadh0415} (\bibinfo {year} {2023})},\
  \Eprint
  {https://arxiv.org/abs/https://www.science.org/doi/pdf/10.1126/sciadv.adh0415}
  {https://www.science.org/doi/pdf/10.1126/sciadv.adh0415} \BibitemShut
  {NoStop}%
\bibitem [{\citenamefont {Bounds}\ \emph {et~al.}(2024)\citenamefont {Bounds},
  \citenamefont {Duff}, \citenamefont {Tritt}, \citenamefont {Taylor},
  \citenamefont {Coe}, \citenamefont {White},\ and\ \citenamefont
  {Turner}}]{Bounds2024}%
  \BibitemOpen
  \bibfield  {author} {\bibinfo {author} {\bibfnamefont {C.~C.}\ \bibnamefont
  {Bounds}}, \bibinfo {author} {\bibfnamefont {J.~P.}\ \bibnamefont {Duff}},
  \bibinfo {author} {\bibfnamefont {A.}~\bibnamefont {Tritt}}, \bibinfo
  {author} {\bibfnamefont {H.~A.~M.}\ \bibnamefont {Taylor}}, \bibinfo {author}
  {\bibfnamefont {G.~X.}\ \bibnamefont {Coe}}, \bibinfo {author} {\bibfnamefont
  {S.~J.}\ \bibnamefont {White}},\ and\ \bibinfo {author} {\bibfnamefont
  {L.~D.}\ \bibnamefont {Turner}},\ }\bibfield  {title} {\enquote {\bibinfo
  {title} {Quantum spectral analysis by continuous measurement of
  {L}andau-{Z}ener transitions},}\ }\href
  {https://doi.org/10.1103/PhysRevLett.132.093401} {\bibfield  {journal}
  {\bibinfo  {journal} {Phys. Rev. Lett.}\ }\textbf {\bibinfo {volume} {132}},\
  \bibinfo {pages} {093401} (\bibinfo {year} {2024})}\BibitemShut {NoStop}%
\bibitem [{\citenamefont {Baake}, \citenamefont {Baake},\ and\ \citenamefont
  {Wagner}(1997)}]{Baake1997}%
  \BibitemOpen
  \bibfield  {author} {\bibinfo {author} {\bibfnamefont {E.}~\bibnamefont
  {Baake}}, \bibinfo {author} {\bibfnamefont {M.}~\bibnamefont {Baake}},\ and\
  \bibinfo {author} {\bibfnamefont {H.}~\bibnamefont {Wagner}},\ }\bibfield
  {title} {\enquote {\bibinfo {title} {Ising quantum chain is equivalent to a
  model of biological evolution},}\ }\href
  {https://doi.org/10.1103/PhysRevLett.78.559} {\bibfield  {journal} {\bibinfo
  {journal} {Phys. Rev. Lett.}\ }\textbf {\bibinfo {volume} {78}},\ \bibinfo
  {pages} {559--562} (\bibinfo {year} {1997})}\BibitemShut {NoStop}%
\bibitem [{\citenamefont {Dziarmaga}(2005)}]{Dziarmaga2005}%
  \BibitemOpen
  \bibfield  {author} {\bibinfo {author} {\bibfnamefont {J.}~\bibnamefont
  {Dziarmaga}},\ }\bibfield  {title} {\enquote {\bibinfo {title} {Dynamics of a
  quantum phase transition: Exact solution of the quantum {I}sing model},}\
  }\href {https://doi.org/10.1103/PhysRevLett.95.245701} {\bibfield  {journal}
  {\bibinfo  {journal} {Phys. Rev. Lett.}\ }\textbf {\bibinfo {volume} {95}},\
  \bibinfo {pages} {245701} (\bibinfo {year} {2005})}\BibitemShut {NoStop}%
\bibitem [{\citenamefont {Kenmoe}\ \emph {et~al.}(2013)\citenamefont {Kenmoe},
  \citenamefont {Phien}, \citenamefont {Kiselev},\ and\ \citenamefont
  {Fai}}]{Kenmoe2013}%
  \BibitemOpen
  \bibfield  {author} {\bibinfo {author} {\bibfnamefont {M.~B.}\ \bibnamefont
  {Kenmoe}}, \bibinfo {author} {\bibfnamefont {H.~O.}\ \bibnamefont {Phien}},
  \bibinfo {author} {\bibfnamefont {M.~N.}\ \bibnamefont {Kiselev}},\ and\
  \bibinfo {author} {\bibfnamefont {L.~C.}\ \bibnamefont {Fai}},\ }\bibfield
  {title} {\enquote {\bibinfo {title} {Effects of colored noise on
  {L}andau-{Z}ener transitions: Two- and three-level systems},}\ }\href
  {https://doi.org/10.1103/PhysRevB.87.224301} {\bibfield  {journal} {\bibinfo
  {journal} {Helv. Phys. Act}\ }\textbf {\bibinfo {volume} {87}},\ \bibinfo
  {pages} {224301} (\bibinfo {year} {2013})}\BibitemShut {NoStop}%
\bibitem [{\citenamefont {Kholodenko}\ and\ \citenamefont
  {Silagadze}(2012)}]{Kholodenko2012}%
  \BibitemOpen
  \bibfield  {author} {\bibinfo {author} {\bibfnamefont {A.~L.}\ \bibnamefont
  {Kholodenko}}\ and\ \bibinfo {author} {\bibfnamefont {Z.~K.}\ \bibnamefont
  {Silagadze}},\ }\bibfield  {title} {\enquote {\bibinfo {title} {When physics
  helps mathematics: {C}alculation of the sophisticated multiple integral},}\
  }\href {https://doi.org/10.1134/S1063779612060068} {\bibfield  {journal}
  {\bibinfo  {journal} {Physics of Particles and Nuclei}\ }\textbf {\bibinfo
  {volume} {43}},\ \bibinfo {pages} {882--888} (\bibinfo {year}
  {2012})}\BibitemShut {NoStop}%
\bibitem [{\citenamefont {Silagadze}(2013)}]{Siladadze2013}%
  \BibitemOpen
  \bibfield  {author} {\bibinfo {author} {\bibfnamefont {Z.~K.}\ \bibnamefont
  {Silagadze}},\ }\bibfield  {title} {\enquote {\bibinfo {title} {Three ways to
  calculate ${I}_2$ ({AMM} problem 11621)},}\ }\href
  {https://wwwsnd.inp.nsk.su/~silagadz/I2.pdf} {\bibfield  {journal} {\bibinfo
  {journal} {notes}\ } (\bibinfo {year} {2013})}\BibitemShut {NoStop}%
\bibitem [{\citenamefont {Feldbrugge}, \citenamefont {Pen},\ and\ \citenamefont
  {Turok}(2023)}]{Job2023}%
  \BibitemOpen
  \bibfield  {author} {\bibinfo {author} {\bibfnamefont {J.}~\bibnamefont
  {Feldbrugge}}, \bibinfo {author} {\bibfnamefont {U.-L.}\ \bibnamefont
  {Pen}},\ and\ \bibinfo {author} {\bibfnamefont {N.}~\bibnamefont {Turok}},\
  }\bibfield  {title} {\enquote {\bibinfo {title} {Oscillatory path integrals
  for radio astronomy},}\ }\href
  {https://doi.org/https://doi.org/10.1016/j.aop.2023.169255} {\bibfield
  {journal} {\bibinfo  {journal} {Annals of Physics}\ }\textbf {\bibinfo
  {volume} {451}},\ \bibinfo {pages} {169255} (\bibinfo {year}
  {2023})}\BibitemShut {NoStop}%
\bibitem [{\citenamefont {Erd{\'e}lyi}(1956)}]{Erdelyi1956}%
  \BibitemOpen
  \bibfield  {author} {\bibinfo {author} {\bibfnamefont {A.}~\bibnamefont
  {Erd{\'e}lyi}},\ }\href {https://books.google.de/books?id=aedk-OHdmNYC}
  {\emph {\bibinfo {title} {Asymptotic Expansions}}},\ Dover Books on
  Mathematics\ (\bibinfo  {publisher} {Dover Publications},\ \bibinfo {year}
  {1956})\BibitemShut {NoStop}%
\bibitem [{\citenamefont {Chen}\ and\ \citenamefont {Bremer}(2024)}]{Chen2024}%
  \BibitemOpen
  \bibfield  {author} {\bibinfo {author} {\bibfnamefont {S.~K.}\ \bibnamefont
  {Chen}, \bibfnamefont {Shukui}}\ and\ \bibinfo {author} {\bibfnamefont
  {J.}~\bibnamefont {Bremer}},\ }\bibfield  {title} {\enquote {\bibinfo {title}
  {On the adaptive levin method},}\ }\href
  {https://doi.org/10.1007/s00211-024-01443-6} {\bibfield  {journal} {\bibinfo
  {journal} {Numerische Mathematik}\ }\textbf {\bibinfo {volume} {156}},\
  \bibinfo {pages} {1927--1985} (\bibinfo {year} {2024})}\BibitemShut {NoStop}%
\bibitem [{\citenamefont {Liu}, \citenamefont {Tian},\ and\ \citenamefont
  {You}(2017)}]{Liu2017}%
  \BibitemOpen
  \bibfield  {author} {\bibinfo {author} {\bibfnamefont {Z.}~\bibnamefont
  {Liu}}, \bibinfo {author} {\bibfnamefont {H.}~\bibnamefont {Tian}},\ and\
  \bibinfo {author} {\bibfnamefont {X.}~\bibnamefont {You}},\ }\bibfield
  {title} {\enquote {\bibinfo {title} {Adiabatic filon-type methods for highly
  oscillatory second-order ordinary differential equations},}\ }\href
  {https://doi.org/https://doi.org/10.1016/j.cam.2017.01.028} {\bibfield
  {journal} {\bibinfo  {journal} {Journal of Computational and Applied
  Mathematics}\ }\textbf {\bibinfo {volume} {320}},\ \bibinfo {pages} {1--14}
  (\bibinfo {year} {2017})}\BibitemShut {NoStop}%
\bibitem [{\citenamefont {Gao}(2021)}]{Jing2012}%
  \BibitemOpen
  \bibfield  {author} {\bibinfo {author} {\bibfnamefont {J.}~\bibnamefont
  {Gao}},\ }\bibfield  {title} {\enquote {\bibinfo {title} {Asymptotic
  expansion of the integral with two oscillations on an infinite interval},}\
  }\href {https://doi.org/https://doi.org/10.1016/j.na.2021.112503} {\bibfield
  {journal} {\bibinfo  {journal} {Nonlinear Analysis}\ }\textbf {\bibinfo
  {volume} {213}},\ \bibinfo {pages} {112503} (\bibinfo {year}
  {2021})}\BibitemShut {NoStop}%
\bibitem [{\citenamefont {Landau}(1932)}]{Landau1932}%
  \BibitemOpen
  \bibfield  {author} {\bibinfo {author} {\bibfnamefont {L.~D.}\ \bibnamefont
  {Landau}},\ }\bibfield  {title} {\enquote {\bibinfo {title} {On the theory of
  transfer of energy at collisions ii},}\ }\href@noop {} {\bibfield  {journal}
  {\bibinfo  {journal} {Phys. Z. Sowjetunion}\ }\textbf {\bibinfo {volume}
  {2}},\ \bibinfo {pages} {46} (\bibinfo {year} {1932})}\BibitemShut {NoStop}%
\bibitem [{\citenamefont {Zener}(1932)}]{Zener1932}%
  \BibitemOpen
  \bibfield  {author} {\bibinfo {author} {\bibfnamefont {C.}~\bibnamefont
  {Zener}},\ }\bibfield  {title} {\enquote {\bibinfo {title} {Non-adiabatic
  crossing of energy levels},}\ }\href {https://doi.org/10.1098/rspa.1932.0165}
  {\bibfield  {journal} {\bibinfo  {journal} {Proc. R. Soc. Lond. A}\ ,\
  \bibinfo {pages} {137696–702}} (\bibinfo {year} {1932})}\BibitemShut
  {NoStop}%
\bibitem [{\citenamefont {Stuckelberg}(1932)}]{Stuckelberg1932}%
  \BibitemOpen
  \bibfield  {author} {\bibinfo {author} {\bibfnamefont {E.~C.~G.}\
  \bibnamefont {Stuckelberg}},\ }\href@noop {} {\bibfield  {journal} {\bibinfo
  {journal} {Helv. Phys. Acta}\ }\textbf {\bibinfo {volume} {5}},\ \bibinfo
  {pages} {369} (\bibinfo {year} {1932})}\BibitemShut {NoStop}%
\bibitem [{\citenamefont {Majorana}(1932)}]{Majorana1932}%
  \BibitemOpen
  \bibfield  {author} {\bibinfo {author} {\bibfnamefont {E.}~\bibnamefont
  {Majorana}},\ }\bibfield  {title} {\enquote {\bibinfo {title} {Atomi
  orientati in campo magnetico variabile},}\ }\href
  {https://doi.org/10.1007/BF02960953} {\bibfield  {journal} {\bibinfo
  {journal} {Nuovo Cimento (1924-1942)}\ }\textbf {\bibinfo {volume} {9}},\
  \bibinfo {pages} {1827--6121} (\bibinfo {year} {1932})}\BibitemShut {NoStop}%
\bibitem [{\citenamefont {Bloch}\ and\ \citenamefont {Rojo}(2010)}]{Rojo2010a}%
  \BibitemOpen
  \bibfield  {author} {\bibinfo {author} {\bibfnamefont {A.~M.}\ \bibnamefont
  {Bloch}}\ and\ \bibinfo {author} {\bibfnamefont {A.~G.}\ \bibnamefont
  {Rojo}},\ }\bibfield  {title} {\enquote {\bibinfo {title} {Kinematics of the
  rolling sphere and quantum spin},}\ }\href
  {https://doi.org/10.4310/CIS.2010.v10.n4.a4} {\bibfield  {journal} {\bibinfo
  {journal} {Communications in Information and Systems}\ }\textbf {\bibinfo
  {volume} {10}},\ \bibinfo {pages} {221--238} (\bibinfo {year}
  {2010})}\BibitemShut {NoStop}%
\bibitem [{\citenamefont {Rojo}\ and\ \citenamefont {Bloch}(2010)}]{Rojo2010b}%
  \BibitemOpen
  \bibfield  {author} {\bibinfo {author} {\bibfnamefont {A.~G.}\ \bibnamefont
  {Rojo}}\ and\ \bibinfo {author} {\bibfnamefont {A.~M.}\ \bibnamefont
  {Bloch}},\ }\bibfield  {title} {\enquote {\bibinfo {title} {The rolling
  sphere, the quantum spin, and a simple view of the landau–zener problem},}\
  }\href {https://doi.org/10.1119/1.3456565} {\bibfield  {journal} {\bibinfo
  {journal} {American Journal of Physics}\ }\textbf {\bibinfo {volume} {78}},\
  \bibinfo {pages} {1014--1022} (\bibinfo {year} {2010})},\ \Eprint
  {https://arxiv.org/abs/https://pubs.aip.org/aapt/ajp/article-pdf/78/10/1014/13086269/1014\_1\_online.pdf}
  {https://pubs.aip.org/aapt/ajp/article-pdf/78/10/1014/13086269/1014\_1\_online.pdf}
  \BibitemShut {NoStop}%
\bibitem [{\citenamefont {Beck}(2025)}]{Dominik2025}%
  \BibitemOpen
  \bibfield  {author} {\bibinfo {author} {\bibfnamefont {D.}~\bibnamefont
  {Beck}},\ }\bibfield  {title} {\enquote {\bibinfo {title} {When mathematics
  helps physics: Calculation of the integral of kholodenko and silagadze},}\
  }\href {https://doi.org/https://doi.org/10.48550/arXiv.2504.00134} {\bibfield
   {journal} {\bibinfo  {journal} {ArXiv}\ } (\bibinfo {year} {2025}),\
  https://doi.org/10.48550/arXiv.2504.00134}\BibitemShut {NoStop}%
\bibitem [{\citenamefont {Abramowitz}\ and\ \citenamefont
  {Stegun}(1965)}]{Abramowitz}%
  \BibitemOpen
  \bibinfo {editor} {\bibfnamefont {M.}~\bibnamefont {Abramowitz}}\ and\
  \bibinfo {editor} {\bibfnamefont {I.~A.}\ \bibnamefont {Stegun}},\ eds.,\
  \href@noop {} {\emph {\bibinfo {title} {Handbook of Mathematical Functions
  with Formulas, Graphs and Mathematical Tables}}}\ (\bibinfo  {publisher}
  {Dover Publications, Inc.},\ \bibinfo {address} {New York},\ \bibinfo {year}
  {1965})\BibitemShut {NoStop}%
\bibitem [{\citenamefont {Dunkl}, \citenamefont {Ismail},\ and\ \citenamefont
  {Wong}(2000)}]{Wong2000}%
  \BibitemOpen
  \bibfield  {author} {\bibinfo {author} {\bibfnamefont {C.}~\bibnamefont
  {Dunkl}}, \bibinfo {author} {\bibfnamefont {M.}~\bibnamefont {Ismail}},\ and\
  \bibinfo {author} {\bibfnamefont {R.}~\bibnamefont {Wong}},\ }\href
  {https://doi.org/10.1142/4502} {\emph {\bibinfo {title} {Special
  Functions}}}\ (\bibinfo  {publisher} {WORLD SCIENTIFIC},\ \bibinfo {year}
  {2000})\ \Eprint
  {https://arxiv.org/abs/https://www.worldscientific.com/doi/pdf/10.1142/4502}
  {https://www.worldscientific.com/doi/pdf/10.1142/4502} \BibitemShut {NoStop}%
\bibitem [{\citenamefont {Fai}(2024)}]{Fai2024}%
  \BibitemOpen
  \bibfield  {author} {\bibinfo {author} {\bibfnamefont {L.~C.}\ \bibnamefont
  {Fai}},\ }\href {https://doi.org/10.1088/978-0-7503-6149-1} {\emph {\bibinfo
  {title} {Special Functions in Physics and Engineering}}},\ 2053-2563\
  (\bibinfo  {publisher} {IOP Publishing},\ \bibinfo {year} {2024})\BibitemShut
  {NoStop}%
\bibitem [{\citenamefont {Nyisomeh}, \citenamefont {Ateuafack},\ and\
  \citenamefont {Fai}(2019)}]{Nyisomeh2019}%
  \BibitemOpen
  \bibfield  {author} {\bibinfo {author} {\bibfnamefont {I.}~\bibnamefont
  {Nyisomeh}}, \bibinfo {author} {\bibfnamefont {M.}~\bibnamefont
  {Ateuafack}},\ and\ \bibinfo {author} {\bibfnamefont {L.}~\bibnamefont
  {Fai}},\ }\bibfield  {title} {\enquote {\bibinfo {title} {Noise-induced
  multilevel {L}andau-{Z}ener transitions: Density matrix investigation},}\
  }\href {https://doi.org/https://doi.org/10.1016/j.physleta.2019.01.035}
  {\bibfield  {journal} {\bibinfo  {journal} {Physics Letters A}\ }\textbf
  {\bibinfo {volume} {383}},\ \bibinfo {pages} {1350--1356} (\bibinfo {year}
  {2019})}\BibitemShut {NoStop}%
\bibitem [{\citenamefont {Luo}\ and\ \citenamefont {Raikh}(2017)}]{Luo2017}%
  \BibitemOpen
  \bibfield  {author} {\bibinfo {author} {\bibfnamefont {Z.-X.}\ \bibnamefont
  {Luo}}\ and\ \bibinfo {author} {\bibfnamefont {M.~E.}\ \bibnamefont
  {Raikh}},\ }\bibfield  {title} {\enquote {\bibinfo {title} {{L}andau-{Z}ener
  transition driven by slow noise},}\ }\href
  {https://doi.org/10.1103/PhysRevB.95.064305} {\bibfield  {journal} {\bibinfo
  {journal} {Phys. Rev. B}\ }\textbf {\bibinfo {volume} {95}},\ \bibinfo
  {pages} {064305} (\bibinfo {year} {2017})}\BibitemShut {NoStop}%
\bibitem [{\citenamefont {Garanin}\ and\ \citenamefont
  {Schilling}(2005)}]{Garanin2005}%
  \BibitemOpen
  \bibfield  {author} {\bibinfo {author} {\bibfnamefont {D.~A.}\ \bibnamefont
  {Garanin}}\ and\ \bibinfo {author} {\bibfnamefont {R.}~\bibnamefont
  {Schilling}},\ }\bibfield  {title} {\enquote {\bibinfo {title} {Many-body
  {L}andau-{Z}ener effect at fast sweep},}\ }\href
  {https://doi.org/10.1103/PhysRevB.71.184414} {\bibfield  {journal} {\bibinfo
  {journal} {Phys. Rev. B}\ }\textbf {\bibinfo {volume} {71}},\ \bibinfo
  {pages} {184414} (\bibinfo {year} {2005})}\BibitemShut {NoStop}%
\bibitem [{\citenamefont {Enomoto}\ and\ \citenamefont
  {Matsuda}(2022)}]{Enomoto2022}%
  \BibitemOpen
  \bibfield  {author} {\bibinfo {author} {\bibfnamefont {S.}~\bibnamefont
  {Enomoto}}\ and\ \bibinfo {author} {\bibfnamefont {T.}~\bibnamefont
  {Matsuda}},\ }\bibfield  {title} {\enquote {\bibinfo {title} {The exact {WKB}
  and the {L}andau-{Z}ener transition for asymmetry in cosmological particle
  production},}\ }\href {https://doi.org/10.1007/JHEP02(2022)131} {\bibfield
  {journal} {\bibinfo  {journal} {Journal of High Energy Physics}\ }\textbf
  {\bibinfo {volume} {2022}},\ \bibinfo {pages} {1029--8479} (\bibinfo {year}
  {2022})}\BibitemShut {NoStop}%
\bibitem [{\citenamefont {Damski}(2005)}]{Damski2005}%
  \BibitemOpen
  \bibfield  {author} {\bibinfo {author} {\bibfnamefont {B.}~\bibnamefont
  {Damski}},\ }\bibfield  {title} {\enquote {\bibinfo {title} {The simplest
  quantum model supporting the {K}ibble-{Z}urek mechanism of topological defect
  production: {L}andau-{Z}ener transitions from a new perspective},}\ }\href
  {https://doi.org/10.1103/PhysRevLett.95.035701} {\bibfield  {journal}
  {\bibinfo  {journal} {Phys. Rev. Lett.}\ }\textbf {\bibinfo {volume} {95}},\
  \bibinfo {pages} {035701} (\bibinfo {year} {2005})}\BibitemShut {NoStop}%
\bibitem [{\citenamefont {Kou}, \citenamefont {Zhang},\ and\ \citenamefont
  {Li}(2025)}]{Kou2025}%
  \BibitemOpen
  \bibfield  {author} {\bibinfo {author} {\bibfnamefont {H.-C.}\ \bibnamefont
  {Kou}}, \bibinfo {author} {\bibfnamefont {Z.-H.}\ \bibnamefont {Zhang}},\
  and\ \bibinfo {author} {\bibfnamefont {P.}~\bibnamefont {Li}},\ }\bibfield
  {title} {\enquote {\bibinfo {title} {{K}ibble-{Z}urek scaling immune to
  anti-{K}ibble-{Z}urek behavior in driven open systems at the limit of loss
  difference},}\ }\href {https://doi.org/10.1103/PhysRevB.111.155152}
  {\bibfield  {journal} {\bibinfo  {journal} {Phys. Rev. B}\ }\textbf {\bibinfo
  {volume} {111}},\ \bibinfo {pages} {155152} (\bibinfo {year}
  {2025})}\BibitemShut {NoStop}%
\bibitem [{\citenamefont {Saito}\ \emph {et~al.}(2006)\citenamefont {Saito},
  \citenamefont {Wubs}, \citenamefont {Kohler}, \citenamefont {Hänggi},\ and\
  \citenamefont {Kayanuma}}]{Saito2006}%
  \BibitemOpen
  \bibfield  {author} {\bibinfo {author} {\bibfnamefont {K.}~\bibnamefont
  {Saito}}, \bibinfo {author} {\bibfnamefont {M.}~\bibnamefont {Wubs}},
  \bibinfo {author} {\bibfnamefont {S.}~\bibnamefont {Kohler}}, \bibinfo
  {author} {\bibfnamefont {P.}~\bibnamefont {Hänggi}},\ and\ \bibinfo {author}
  {\bibfnamefont {Y.}~\bibnamefont {Kayanuma}},\ }\bibfield  {title} {\enquote
  {\bibinfo {title} {Quantum state preparation in circuit {QED} via
  {L}andau-{Z}ener tunneling},}\ }\href
  {https://doi.org/10.1209/epl/i2006-10232-4} {\bibfield  {journal} {\bibinfo
  {journal} {Europhysics Letters}\ }\textbf {\bibinfo {volume} {76}},\ \bibinfo
  {pages} {22} (\bibinfo {year} {2006})}\BibitemShut {NoStop}%
\bibitem [{\citenamefont {Vitanov}(1999)}]{Vitanov1999}%
  \BibitemOpen
  \bibfield  {author} {\bibinfo {author} {\bibfnamefont {N.~V.}\ \bibnamefont
  {Vitanov}},\ }\bibfield  {title} {\enquote {\bibinfo {title} {Transition
  times in the {L}andau-{Z}ener model},}\ }\href
  {https://doi.org/10.1103/PhysRevA.59.988} {\bibfield  {journal} {\bibinfo
  {journal} {Phys. Rev. A}\ }\textbf {\bibinfo {volume} {59}},\ \bibinfo
  {pages} {988--994} (\bibinfo {year} {1999})}\BibitemShut {NoStop}%
\bibitem [{Sil(2012)}]{Silagadze2012}%
  \BibitemOpen
  \bibfield  {title} {\enquote {\bibinfo {title} {Problems and solutions},}\
  }\href {https://www.jstor.org/stable/10.4169/amer.math.monthly.119.02.161}
  {\bibfield  {journal} {\bibinfo  {journal} {The American Mathematical
  Monthly}\ }\textbf {\bibinfo {volume} {119}},\ \bibinfo {pages} {pp.
  161--168} (\bibinfo {year} {2012})}\BibitemShut {NoStop}%
\bibitem [{\citenamefont {de~Reyna}(2015)}]{Juan2015}%
  \BibitemOpen
  \bibfield  {author} {\bibinfo {author} {\bibfnamefont {J.~A.}\ \bibnamefont
  {de~Reyna}},\ }\bibfield  {title} {\enquote {\bibinfo {title} {Cancellations
  in power series of sine type},}\ }\href
  {https://doi.org/10.48550/arXiv.1505.00440} {\bibfield  {journal} {\bibinfo
  {journal} {ArXiv}\ } (\bibinfo {year} {2015})}\BibitemShut {NoStop}%
\bibitem [{\citenamefont {Mathar}(2024)}]{Richard2014}%
  \BibitemOpen
  \bibfield  {author} {\bibinfo {author} {\bibfnamefont {R.~J.}\ \bibnamefont
  {Mathar}},\ }\bibfield  {title} {\enquote {\bibinfo {title} {Yet another
  table of integrals},}\ }\href {https://doi.org/10.48550/arXiv.1207.5845}
  {\bibfield  {journal} {\bibinfo  {journal} {ArXiv}\ } (\bibinfo {year}
  {2024})}\BibitemShut {NoStop}%
\bibitem [{\citenamefont {Comtet}(1974)}]{comtet1974}%
  \BibitemOpen
  \bibfield  {author} {\bibinfo {author} {\bibfnamefont {L.}~\bibnamefont
  {Comtet}},\ }\href {https://doi.org/10.1007/978-94-010-2196-8} {\emph
  {\bibinfo {title} {Advanced Combinatorics: The Art of Finite and Infinite
  Expansions}}}\ (\bibinfo  {publisher} {Reidel Publishing Company},\ \bibinfo
  {address} {Dordrecht, Holland},\ \bibinfo {year} {1974})\ \bibinfo {note}
  {translated from the French by J. W. Nienhuys}\BibitemShut {NoStop}%
\bibitem [{\citenamefont {Vitanov}\ and\ \citenamefont
  {Garraway}(1996)}]{Vitanov1996}%
  \BibitemOpen
  \bibfield  {author} {\bibinfo {author} {\bibfnamefont {N.~V.}\ \bibnamefont
  {Vitanov}}\ and\ \bibinfo {author} {\bibfnamefont {B.~M.}\ \bibnamefont
  {Garraway}},\ }\bibfield  {title} {\enquote {\bibinfo {title}
  {{L}andau-{Z}ener model: Effects of finite coupling duration},}\ }\href
  {https://doi.org/10.1103/PhysRevA.53.4288} {\bibfield  {journal} {\bibinfo
  {journal} {Phys. Rev. A}\ }\textbf {\bibinfo {volume} {53}},\ \bibinfo
  {pages} {4288--4304} (\bibinfo {year} {1996})}\BibitemShut {NoStop}%
\bibitem [{\citenamefont {Wolfram~Research}(2019)}]{Mathematica}%
  \BibitemOpen
  \bibfield  {author} {\bibinfo {author} {\bibfnamefont {I.}~\bibnamefont
  {Wolfram~Research}},\ }\enquote {\bibinfo {title} {Mathematica},}\ \
  (\bibinfo  {publisher} {Wolfram Research, Inc.},\ \bibinfo {address}
  {Champaign, Illinois},\ \bibinfo {year} {2019})\BibitemShut {NoStop}%
\bibitem [{\citenamefont {Yan}\ and\ \citenamefont {Wu}(2010)}]{Yan}%
  \BibitemOpen
  \bibfield  {author} {\bibinfo {author} {\bibfnamefont {Y.}~\bibnamefont
  {Yan}}\ and\ \bibinfo {author} {\bibfnamefont {B.}~\bibnamefont {Wu}},\
  }\bibfield  {title} {\enquote {\bibinfo {title} {Integral definition of
  transition time in the {L}andau-{Z}ener model},}\ }\href
  {https://doi.org/10.1103/PhysRevA.81.022126} {\bibfield  {journal} {\bibinfo
  {journal} {Phys. Rev. A}\ }\textbf {\bibinfo {volume} {81}},\ \bibinfo
  {pages} {022126} (\bibinfo {year} {2010})}\BibitemShut {NoStop}%
\bibitem [{\citenamefont {Brataas}\ and\ \citenamefont
  {Rashba}(2011)}]{Brataas2011}%
  \BibitemOpen
  \bibfield  {author} {\bibinfo {author} {\bibfnamefont {A.}~\bibnamefont
  {Brataas}}\ and\ \bibinfo {author} {\bibfnamefont {E.~I.}\ \bibnamefont
  {Rashba}},\ }\bibfield  {title} {\enquote {\bibinfo {title} {Nuclear dynamics
  during {L}andau-{Z}ener singlet-triplet transitions in double quantum
  dots},}\ }\href {https://doi.org/10.1103/PhysRevB.84.045301} {\bibfield
  {journal} {\bibinfo  {journal} {Phys. Rev. B}\ }\textbf {\bibinfo {volume}
  {84}},\ \bibinfo {pages} {045301} (\bibinfo {year} {2011})}\BibitemShut
  {NoStop}%
\bibitem [{\citenamefont {Kovaleva}, \citenamefont {Manevitch},\ and\
  \citenamefont {Kosevich}(2011)}]{Kovaleva}%
  \BibitemOpen
  \bibfield  {author} {\bibinfo {author} {\bibfnamefont {A.}~\bibnamefont
  {Kovaleva}}, \bibinfo {author} {\bibfnamefont {L.~I.}\ \bibnamefont
  {Manevitch}},\ and\ \bibinfo {author} {\bibfnamefont {Y.~A.}\ \bibnamefont
  {Kosevich}},\ }\bibfield  {title} {\enquote {\bibinfo {title} {Fresnel
  integrals and irreversible energy transfer in an oscillatory system with
  time-dependent parameters},}\ }\href
  {https://doi.org/10.1103/PhysRevE.83.026602} {\bibfield  {journal} {\bibinfo
  {journal} {Phys. Rev. E}\ }\textbf {\bibinfo {volume} {83}},\ \bibinfo
  {pages} {026602} (\bibinfo {year} {2011})}\BibitemShut {NoStop}%
\end{thebibliography}%
	\end{document}